%% file: dqHD2arXiv.tex
\documentclass{vldb}

\usepackage{floatrow}
\newfloatcommand{capbtabbox}{table}[][\FBwidth]

\usepackage{blindtext}

\usepackage{PageSetting}
\usepackage{wrapfig,epsfig}

\usepackage{graphicx}

\usepackage[linesnumbered,ruled,vlined]{algorithm2e}
\usepackage{amsmath}
\usepackage{amssymb}
\usepackage{url}

\usepackage{pdfpages}
\usepackage{multirow}

\usepackage{PageSetting}
\usepackage{wrapfig,epsfig}
\usepackage{balance}

\usepackage[linesnumbered,ruled,vlined]{algorithm2e}
\usepackage{amsmath}
\usepackage{amssymb}
\usepackage{url}
\usepackage{enumitem}

\usepackage{times}
\def\done{\hspace*{\fill} $\framebox[2mm]{}$ \medskip}

\title{Hop Doubling Label Indexing for Point-to-Point Distance Querying on Scale-Free Networks}

\numberofauthors{3} 
\author{
Minhao Jiang$^\dag$, Ada Wai-Chee Fu$^\ddag$, Raymond Chi-Wing Wong$^\dag$, Yanyan Xu$^\ddag$\\
\begin{tabular}{ c c }
\affaddr{$^\dag$The Hong Kong University of Science and Technology} & \affaddr{$^\ddag$The Chinese University of Hong Kong} \\
\affaddr{\{mjiangac, raywong\}@cse.ust.hk} & \affaddr{\{adafu, yyxu\}@cse.cuhk.edu.hk}
\end{tabular}
}

\date{\today}

\begin{document}

\maketitle

\begin{sloppy}

\input{abstract}





\input{Intro}

\input{intro2}

\input{HCdimension3}

\input{alg}

\input{alg1}

\input{properties}

\input{4rules}
\input{logD}

\input{pruning}

\input{complexityM2}

\input{Stepping}

\input{BitParallel}


\input{exp3M21}

\input{conclusion}

\vspace*{3mm}

\textbf{ACKNOWLEDGEMENTS}:
We thank the authors of \cite{Akiba13sigmod}
for the PLL coding, and the authors of \cite{JinRXL12sigmod}
for the HCL coding.
We are grateful for the data collections
from SNAP and KONECT. We thank James Cheng for suggestions on the presentation
and responses to reviews for our submission to another conference.
This research was supported by the RGC GRF research grant 412313 Proj\_id 2150758 of Hong Kong.
%

\bibliographystyle{abbrv}

\vspace*{2mm}

\bibliography{ref,ref_ppsp,ref_vcIndex}

\end{sloppy}
\end{document}

%% file: abstract.tex
\begin{abstract}

We study the problem of point-to-point distance querying for massive scale-free
graphs, which is important for numerous applications. Given a directed or undirected graph, we propose to build an index for
answering such queries based on a
hop-doubling labeling technique. We derive bounds on the index size,
the computation costs and I/O costs based on the
properties of unweighted scale-free graphs. We show that our method
is much more efficient compared to the state-of-the-art technique, in terms of both
querying time and indexing time.
Our empirical study shows that our method can handle graphs that are
orders of magnitude larger than existing methods.
\end{abstract} 

%% file: Intro.tex
\section{Introduction}
\label{sec:Intro}

We study the problem of point-to-point distance querying for massive scale-free networks or graphs.
Given a scale-free graph $G=(V,E)$, we aim to answer queries about the distance of a shortest path from a vertex $s$ to a vertex $t$ in the graph.
Such querying is a basic building block in the solutions of many
practical problems including
page similarity in web graphs,
keyword search on RDF graphs \cite{Kargar11vldb}, and network analysis
such as betweenness centrality computation
\cite{Lee12www}. Indirectly it is useful for community detection and
locating influential users in the network.
We give our problem definition as follows.

{\it Problem Definition}.\label{prob}
Let $G = (V,E)$ be a directed unweighted graph, with vertex set $V$ and edge set $E$.
Each edge $(u,v) \in E$ has a
length of $dist_G(u,v) = 1$.
Given an edge $(u,v)$, we say that $v$ is an \emph{out-neighbor} of $u$, and $u$ is an \emph{in-neighbor} of $v$.
A path $p = (v_1, ...,v_l)$ is a sequence of $l$ vertices in $V$ such that for each $v_i ( 1 \leq i < l)$, $(v_i, v_{i+1}) \in E$. (We also denote $p$ by $v_1 \leadsto v_l$.)
The \emph{length} of a path $p$, denoted by $\ell(p)$, is the sum of the lengths
of the edges on $p$. Given $u,v \in V$, the \emph{distance} from $u$ to $v$, denoted by
$dist_G(u,v)$,
is the minimum length of all paths from $u$ to $v$.
If no path $u \leadsto v$ exists, then $dist_G(u,v) = \infty$.
A path $u \leadsto v$ with a length of $dist_G(u,v)$ is a shortest path from $u$ to $v$.
We study the following problem:
\emph{given a static directed unweighted scale-free graph $G = (V, E)$, construct a disk-based index for
processing point-to-point (P2P) distance queries, where a P2P distance
query $dist(s,t)$ is : given
$s, t \in V$, find $dist_G(s,t)$}.

Although distance querying can be readily handled by
Dijkstra's algorithm \cite{Dijkstra59}, the emergence of large
networks such as social networks, RDF graphs, and phone networks
has created new challenges.
The problem of P2P distance querying has been well studied for road networks. Some previous works include \cite{AbrahamDGW11wea,SandersS05esa,GeisbergerSSD08wea,BauerDSSSW10jea,SametSA08sigmod,SankaranarayananSA09pvldb,Tao11sigmod}.
For other graph types, many indexing methods have been proposed. However, the previous works of \cite{ChangYQCQ12vldbj,ChengY09edbt,CohenHKZ03siamcomp,JinRXL12sigmod,SchenkelTW04edbt,Wei10sigmod,
XiaoWPWH09edbt}
 can only handle relatively small graphs due to high index construction cost and large index storage space. For the 2 largest real graphs tested in these studies, we have $|V|$=581K and $|E|/|V|$ = 2.45 \cite{ChangYQCQ12vldbj}, and $|V|$ = 694K and $|E|/|V|$ = 0.45 \cite{JinRXL12sigmod}, respectively.
The more recent works of IS-Label in \cite{Fu13vldb} and the pruned landmark labeling (PLL) scheme in \cite{Akiba13sigmod} can handle bigger graphs. Both are 2-hop labeling methods \cite{CohenHKZ03siamcomp}.

{\it Challenges.}
While the labeling technique has been adopted by the state-of-the-art
indexing algorithms, there are some major challenges related to this technique.
The first challenge is that no existing work has been able to provide a
guarantee of a small label size. The total label size is $O(|V|^2)$ and
 in the worst case, this is the same size as that of a pairwise distance table.
For general graphs, it is shown that there exist graphs $G= (V,E)$
for which any 2-hop labeling index must have a total size
of $\Omega ( |E| |V|^{1/2} ) $ \cite{CohenHKZ03siamcomp}. This high index space complexity will be impractical for large graphs.

The second challenge, which is related to the first, is that
no existing work has been able to give an acceptable bounded complexity on
the computation time and the runtime memory space required for the label
construction. Most existing works are in-memory algorithms and require huge memory
consumption. The only existing work that
has bounded memory consumption is IS-Label \cite{Fu13vldb}.
IS-Label builds a hierarchy from the given graph by extracting at each level an independent vertex set. The remaining graph at each step is augmented with edges to preserve distances among the remaining vertices. Labels are constructed top-down in the hierarchy. The hierarchy need not be completed so that a residual graph $G_k$ may remain in memory and querying is handled by both the labels and a bi-Dijkstra search in $G_k$.
However, IS-Label has no guarantee of a small label size, and also
no guarantee on the scalability of the label construction time.
Another problem of IS-Label is that to limit the number of iterations, $k$, during the label construction, instead of building a full index,
a residual graph $G_k$ is kept in main memory. However,
this is not a pure indexing method since it requires loading $G_k$ before querying, and the size of $G_k$ can be large.

For the existing in-memory algorithms including \cite{CohenHKZ03siamcomp,Wei10sigmod,JinRXL12sigmod,Akiba13sigmod},
the time complexity ranges from $O(|V|^2)$ to $O(|E||V|)$.
%
For the PLL scheme in \cite{Akiba13sigmod},
the actual time performance is much better than the $O(|E||V|)$ bound. However, PLL
is main memory based and is not scalable because of a breadth first search process for every vertex and a pruning process that requires the label index to reside in memory.
Hence, a very large main memory is needed
that not only can hold the input graph but also the entire label index with extra
storage for computation. Using 48GB RAM, the biggest graph reported in \cite{Akiba13sigmod} to be handled by PLL is a little over 1GB in size since the label size is 22GB.
 Except for IS-Label, all of the above algorithms assume that the given graph can fit in memory, which may not be true for massive networks. Hence, scalability remains a major challenge.

We propose a new indexing method for distance querying to meet the above challenges. Our design is based
on the properties of unweighted scale-free graphs, which are prevalent in the real world
\cite{KONECT,Bollobas04com,Faloutsos_SIGCOMM99,Newman01phyrev}.
Important applications such as social networks, web and most of the collected datasets in \cite{KONECT} belong to this type of graphs.
We offer guaranteed complexity bounds
on the label size, the computation costs and I/O costs. With only 4GB RAM, we
 are able to build an index for a graph of 9GB in size, with hundreds of millions of vertices and edges. Our method is based on a novel iterative process which minimizes the label size growth at each iteration, leading to highly effective labeling for the index.

Our main contributions are summarized as follows:
(1) We propose a novel 2-hop labeling indexing method for P2P distance querying on
unweighted directed graphs,
and have developed I/O-efficient algorithms for index construction when the given graph and the index cannot fit in main memory.
(2)
Based on the properties of unweighted scale-free graphs, we derive the following complexity bounds for our index: the index size is $O(h|V|)$, the computational cost is $O( |V| log M ( |V|/M + log |V|))$,
and the I/O cost is $O( |V| log |V|/M \times |V|/B )$, where $h$ is a small constant, $M$ is the memory size and $B$ is the disk block size.
(3)
We verify the performance of our method with experiments on
large real-world scale-free networks.

The paper is organized as follows.
Section \ref{sec2}
discusses the relevant properties of scale-free graphs.
Section \ref{sec:alg} introduces our main algorithm Hop-Doubling.
Section \ref{sec:complexity} describes the I/O-efficient
algorithms.
Section \ref{sec:Stepping} introduces the Hop-Stepping strategy for performance enhancement.
Section \ref{sec:othergraphs} is a discussion about the adaptations to undirected and weighted graphs, and
about the use of our method for general graphs.
We report our empirical study
in Section \ref{sec:exp},
and conclude
in Section \ref{concl}.

%% file: intro2.tex
\vspace*{-3mm}
\section{2-Hop Labeling for Scale-Free Graphs}
\label{sec2}


The 2-hop labeling technique constructs \emph{labels} for vertices, and a distance query for $s,t$ can be answered by looking up the labels of $s$ and $t$ only.
Each \textbf{label} is a set of label entries and each label entry is a pair $(v,d)$ where $v \in V$
and $d$ is a distance value. We say that $v$ is a \textbf{pivot}.
For a directed graph $G = (V,E)$, we create two labels $\mathcal{L}_{in}(v)$ and $\mathcal{L}_{out}(v)$ for each
vertex $v \in V$ so that if $dist_G(s,t) \neq \infty$,
then we can find a pivot $u$ such that
$(u, d_1) \in \mathcal{L}_{out}(s)$, $(u, d_2) \in \mathcal{L}_{in}(t)$ and
$d_1+d_2 = dist_G(s,t)$, and there does not exist any $u'$ such that $(u', d_1') \in \mathcal{L}_{out}(s)$, $(u',d_2') \in \mathcal{L}_{in}(t)$ and $d_1'+d_2' < dist_G(s,t)$. We say that the pair $(s,t)$ is \textbf{covered} by $u$.
Hence, the distance query $dist(s,t)$ can be answered by
looking up $\mathcal{L}_{out}(s)$ and $\mathcal{L}_{in}(t)$
for such a pivot $u$ with the smallest $d_1+d_2$.

The set of labels for all vertices is called a \textbf{2-hop cover}. The complexity of finding a minimum 2-hop cover is shown to be NP-hard \cite{CohenHKZ03siamcomp},
and known approximate algorithms are also very costly \cite{JinRXL12sigmod}. However, in the following discussion, we will show that certain ordering of vertices may give rise to a good 2-hop cover, which sheds some light on this hard problem.


\begin{figure}
\begin{floatrow}
\ffigbox{%
\hspace*{-8mm}
\includegraphics[width=3.0cm]{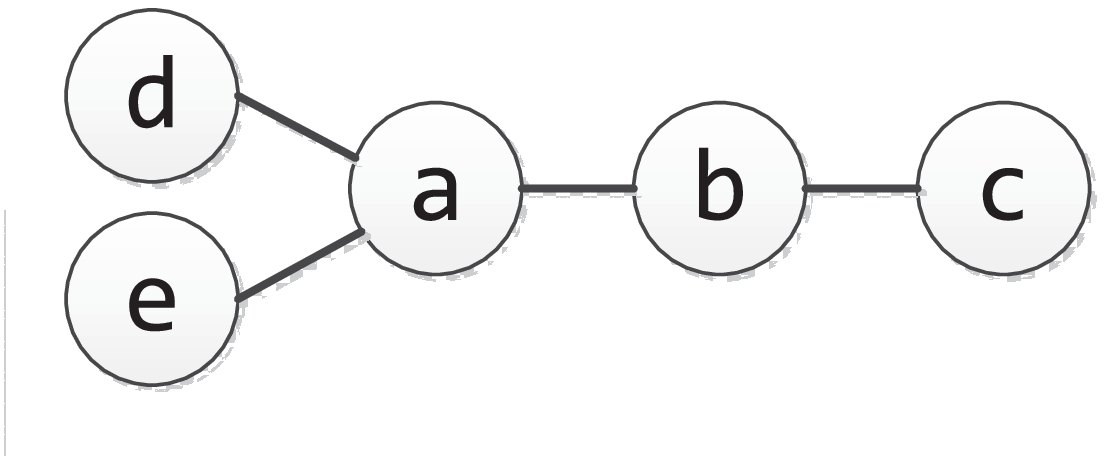}
}{%
\vspace*{-3mm}
\vspace*{-2mm}
  \caption{A road graph $G_R$}%
  \label{fig:roadgraph}
}
\capbtabbox{%
\label{table:roadgraph}
\hspace*{-15mm}
\begin{scriptsize}
\begin{tabular}{|l|l|}
  \hline
  $\mathcal{L}(a)$ & $\{(a,0),(b,1),(c,2), (d,1), (e,1)\}$ \\
  $\mathcal{L}(b)$ & $\{(b,0),(c,1),(d,2),(e,2)\}$ \\
  $\mathcal{L}(c)$ & $\{(c,0),(e,3)\}$ \\
  $\mathcal{L}(d)$ & $\{(d,0),(c,3)\}$\\
  $\mathcal{L}(e)$ & $\{(e,0),(d,2)\}$\\ \hline
\end{tabular}
\end{scriptsize}
}{%
\vspace*{-3mm}
  \caption{A label index for $G_R$}
  \label{table:roadgraph}
}
\end{floatrow}
\end{figure}


\begin{figure}
\begin{floatrow}
\ffigbox{%
\hspace*{-5mm}
\includegraphics[width=2.0cm]{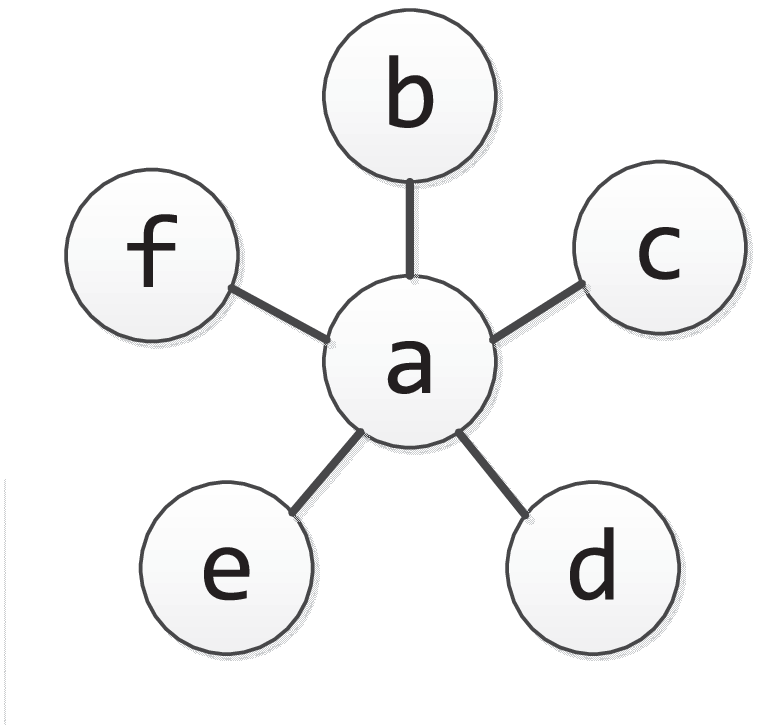}
}{%
\vspace*{-3mm}
  \caption{A star graph $G_S$}%
  \label{fig:stargraph}
}
\capbtabbox{%
\begin{scriptsize}
\hspace*{-8mm}
\begin{tabular}{|l|l|}
  \hline
  $\mathcal{L}(a)$ & $\{(a,0),(b,1),(c,1),(d,1),$\\
  & $(e,1),(f,1)\}$\\
  $\mathcal{L}(b)$ & $\{(b,0),(c,2),(d,2)\}$ \\
  $\mathcal{L}(c)$ & $\{(c,0),(d,2),(e,2)\}$ \\
  $\mathcal{L}(d)$ & $\{(d,0),(e,2),(f,2)\}$\\
  $\mathcal{L}(e)$ & $\{(e,0),(f,2),(b,2)\}$ \\
  $\mathcal{L}(f)$ & $\{(f,0),(b,2),(c,2)\}$\\
  \hline
\end{tabular}
\end{scriptsize}
}{%
\vspace*{-3mm}
  \caption{A label index for $G_S$}
  \label{table:stargraph}
}
\end{floatrow}
%
\vspace*{2mm}
\begin{floatrow}
\capbtabbox{%
\hspace*{-2mm}
\begin{scriptsize}
\begin{tabular}{|l|l|}
  \hline
  $\mathcal{L}(a)$ & $\{(a,0)\}$ \\
  $\mathcal{L}(b)$ & $\{(b,0),(a,1)\}$ \\
  $\mathcal{L}(c)$ & $\{(c,0),(a,2),(b,1)\}$ \\
  $\mathcal{L}(d)$ & $\{(d,0),(a,1)\}$\\
  $\mathcal{L}(e)$ & $\{(e,0),(a,1)\}$ \\
  \hline
\end{tabular}
\vspace*{-3mm}
\end{scriptsize}
}{%
  \hspace*{-3mm}\caption{A small $G_R$ index}
  \label{tab:smallGr}
}
\capbtabbox{%
\hspace*{1mm}
\begin{scriptsize}
\begin{tabular}{|l|l|}
  \hline
  $\mathcal{L}(a)$ & $\{(a,0)\}$ \\
  $\mathcal{L}(b)$ & $\{(b,0),(a,1)\}$ \hspace*{5mm} \\
  $\mathcal{L}(c)$ & $\{(c,0),(a,1)\}$\\
  $\mathcal{L}(d)$ & $\{(d,0),(a,1)\}$\\
  $\mathcal{L}(e)$ & $\{(e,0),(a,1)\}$ \\
  $\mathcal{L}(f)$ & $\{(f,0),(a,1)\}$\\
  \hline
\end{tabular}
\vspace*{-3mm}
\end{scriptsize}
}{%
  \caption{A small $G_S$ index}
  \label{tab:smallGs}
}
\end{floatrow}
\end{figure}




\subsection{Ordering of Vertices for Labeling}
\label{sec:ordering}

The importance of the ordering of vertices can be illustrated by some very simple graphs.
In Figure \ref{fig:roadgraph}, we show a graph $G_R$ for representing a simple road system. $G_R$ is undirected, but we can treat it as directed since each edge can be seen as bidirectional.
Table \ref{table:roadgraph} is a 2-hop cover for $G_R$ where $\mathcal{L}(v) = \mathcal{L}_{in}(v) = \mathcal{L}_{out}(v)$. The 2-hop cover
is \textbf{minimal}, meaning that we cannot delete any label entry and still maintain the correctness of distance query evaluation.
The entries of the form $(v,0)$ are trivial but are needed for query answering.
In Figure \ref{fig:stargraph}, we show a star graph, $G_S$. Table \ref{table:stargraph} is a 2-hop cover for $G_S$ which is also minimal. For example, if we delete $(c,2)$ from $\mathcal{L}(b)$,
then for a query of $dist(b,c)$, we would return an incorrect distance of 4 from $(d,2)$ in $\mathcal{L}(b)$ and $(d,2)$ in $\mathcal{L}(c)$. Note that one can add many useless entries to these covers so that they are still correct but not minimal.





For a given graph, there can be many possible minimal 2-hop covers, and in Tables \ref{tab:smallGr} and \ref{tab:smallGs}, we show smaller minimal 2-hop covers for $G_R$ and $G_S$, which reduce the number of non-trivial label entries by half or more when compared with those shown in Tables \ref{table:roadgraph} and \ref{table:stargraph}. Intuitively, for the road network, we are making use of the hub $a$, which lies on the shortest paths for many pairs of vertices. Similarly, we make heavy use of the center $a$ of the star graph, which has a highest degree. The problem of finding a minimum 2-hop cover is to find a smallest set of label entries with pivots that cover the shortest paths for answering all distance queries, and in these special graphs, the hub or center obviously hits the most number of shortest paths.
We can set a ranking on the vertices in such a way that higher ranked vertices are likely to hit more shortest paths, and then use higher ranked vertices for pivots, as in the examples.
This should result in a smaller label size.

 The above idea is more formally treated by the notion of \emph{canonical labeling} in \cite{AbrahamDGW12esa}. If shortest paths are not unique for given $s,t$, we may define canonical labeling as follows.
Given a total ranking $r()$ of all vertices in $V$, a labeling is canonical if a vertex $v$ is a pivot in $\mathcal{L}_{out}(u)$ if and only if there exists a vertex $w$ such that $r(v)$ is the highest among all vertices in all shortest paths from $u$ to $w$, and similarly for $\mathcal{L}_{in}(u)$. The labeling is minimal since deleting any pivot creates some uncovered pair of vertices. Canonical labeling calls for the \emph{pruning} of any entry $(v,d)$ in $\mathcal{L}_{out}(u)$ if by looking up $\mathcal{L}_{out}(u)$ and $\mathcal{L}_{in}(v)$
we find a higher rank pivot $v'$ that
gives a path $p = ( u, ..., v', ..., v)$ with a length $\leq d$. This is because if $v$ is on a shortest path from $u$ to another vertex $w$ which is made up of $p' = (u, ...,v)$ of length $d$ and
$q = (v, ...,w)$, then
$v'$ will also be on a shortest path from $u$ to $w$, which
is made up of $p$ and $q$. Since
$r(v')>r(v)$, $v$ should not be chosen as a pivot here.

Given the importance of ranking as illustrated in the above examples, 
we expect good indexing results from a good vertex ranking.
The independent set approach of IS-Label \cite{Fu13vldb} effectively gives low ranking to low degree vertices. This ordering is found to produce good label sizes.
The pruned landmark scheme PLL in \cite{Akiba13sigmod} builds labels for an unweighted graph by a breadth first search (BFS) from vertices ordered in non-increasing degrees. The search frontiers of BFS are halted at vertices where the label entries are pruned by previously entered entries as described in the above.
Note that such pruning has also been proposed in \cite{AbrahamDGW12esa}.
This ordering by degree is found to be highly effective
for many real graphs.
In the next subsection, we will derive reasons behind this effectiveness for scale-free graphs.
We make use of the definition of \emph{hitting sets} and a concept similar to the \emph{highway dimension} introduced in \cite{AbrahamFGW10soda,AbrahamDFGW11icalp} for road networks. However, we should point out that the characteristics of a scale-free graph is very much different from that of a road network.

%% file: HCdimension3.tex
\subsection{Hitting Sets in Scale-free Graphs}
\label{sec:scaleFree}

 \label{ssec:hittingSet}



%

A function $f(x)$ is said to be \emph{scale-free} if
$f(bx) = C(b)f(x)$,
where $C(b)$ is some constant dependent only on $b$.
It is common to call a graph scale-free if the distribution of vertex degrees of the graph follows a \emph{power law}:
$ \mbox{Prob(a vertex has degree $k$)} \varpropto k^{-\alpha}$, where $\alpha$ is a positive real number.
%
%
%
%
This is scale-free since if $f(x) = cx^{-\alpha}$, then $f(bx) = c(bx)^{-\alpha} = b^{-\alpha}f(x)$. Typically, $2 \leq \alpha \leq 3$ \cite{Chen12TOA,Bu02infocom,Faloutsos_SIGCOMM99}. Existing works \cite{Bollobas04com,Faloutsos_SIGCOMM99,KONECT,Newman01phyrev} have shown that many real world graphs do follow such power law distributions. Based on the BA model \cite{Barabasi99sci} of scale-free graphs, Bollobas et al. \cite{Bollobas04com} proved that the diameter $D$ of a scale-free random graph is asymptotically
\begin{eqnarray}
\mbox{$D = \log |V|/ \log \log |V|$}
\label{diameter}
\end{eqnarray}
Although this is an asymptotical analysis, it gives very accurate prediction for many real world scale-free graphs \cite{KONECT,Wang03ICSM}.


Newman et al. \cite{Newman01phyrev} studied the properties of scale-free graphs by means of
generating functions for the probability distribution of vertex degrees.
Let $z_i$ be the average number of vertices that are $i$ hops away from
a randomly chosen vertex $v$.
It is shown that with very high probability,
$z_m = (z_2 / z_1)^{m-1} z_1$.
Hence $z_m = (z_2/z_1) z_{m-1}$.
Thus, the \emph{expansion factor} $R$ can be determined by the average number of vertices that are 1 or 2 hops from $v$, respectively, i.e., $R = z_2/z_1$.
With an expansion factor of $R$, the diameter of the graph can be estimated to be
$ D = log_R |V| =  log |V| / log R$.
From Equation (\ref{diameter}), the expansion factor is given by
\begin{eqnarray}
\mbox{$R = log|V|$}
\label{R}
\end{eqnarray}
%

%


For a graph $G=(V,E)$ that follows a power law distribution,
Faloutsos et al. \cite{Faloutsos_SIGCOMM99} derived the following
relationship between the degree $deg_v$ of a vertex $v$ in
$G$ and its
rank in terms of the degree.
For a vertex $v \in V$, $v$ has the $r(v)$-th highest degree in $G$.

\begin{lemma}\cite{Faloutsos_SIGCOMM99}
The degree, $deg_v$, of a vertex $v$, is a function of the rank of the vertex,
$r(v)$, and the rank exponent, $\gamma$, as follows:
\begin{eqnarray}    \label{rank}
deg_v = \frac{1}{|V|^\mathcal{\gamma}} (r(v) ^\mathcal{\gamma})
 \end{eqnarray}
 \label{lemmaRank}
\vspace*{-3mm}
\end{lemma}
In the above,
$\mathcal{\gamma}$ is a small real number
 found
to be between $-0.8$ and $-0.7$ for many real-world graphs
\cite{Faloutsos_SIGCOMM99}.
%
%
According to Equation (\ref{rank}), taking
$\mathcal{\gamma} = -0.8$ for a scale-free graph $G_1 = (V_1, E_1)$,
if $|V_1|$=1M, then less than 500 vertices have degree above 500, and the top-degree vertex $v_0$ has a degree of 63095.
From Equation (\ref{R}), the expansion factor is given by
  $R = \log |V_1| \approx 20$. Since $63095 \times 20 > 1M$,
  $v_0$ is expected to reach all
vertices within 2 hops.


Let us call the number of hops (edges) on
a path its \textbf{hop length}.
Given a set of paths $P$, a \textbf{hitting set} for $P$ is a set of vertices $S$ such that each path $p$ in $P$ contains at least one vertex $v$ in $S$ (we say that $p$ is \textbf{hit} by $v$). 
 For the above graph $G_1$, a single highest degree vertex is expected to
hit all shortest paths with length $\geq 4$. In general, we make an assumption of a \emph{small} hitting set for \emph{long} shortest paths as follows.

\begin{assumption}
Given an unweighted scale-free graph $G=(V,E)$,
there exist small integers $d_0$ and $h$, and
a set $\mathbb{H}$ of the highest degree vertices in $V$, such that
$\forall u,v \in V$, if there exist shortest paths $u \leadsto v$ with
hop length $\geq d_0$, then one such path is hit by one of $h$ vertices in $\mathbb{H}$.
\label{A1}
\end{assumption}

In Assumption \ref{A1}, $|\mathbb{H}| \geq h$.
Given Equations (\ref{diameter}) to (\ref{rank}), we can show that
Assumption \ref{A1} holds with $d_0 =4 $ and $h=1$ for any undirected unweighted scale-free graph $G=(V,E)$ with $|V| \geq 3$, and rank exponent $-0.8 \leq \gamma \leq - 0.7$
(typical values in real world graphs
\cite{Faloutsos_SIGCOMM99}).
%
The analysis goes as follows.
From Lemma \ref{lemmaRank}, the degree of $v_0$ is given by $|V|^{- \gamma}$
since $r(v_0) = 1$. With an expansion factor of $R$,
if $(deg_{v_0} \times R )\geq |V|$, then $v_0$ reaches all vertices in 2 hops.
This is the case where
$({|V|^{-\gamma}} \cdot R ) \geq |V|$,
and from Equation \ref{R}, $R = \log |V|$; hence the inequality 
becomes
{$(|V|^{-\gamma - 1} \cdot \log |V|) \geq 1$},
and this holds for all values of $|V| \geq 3$ for $-0.8 \leq \gamma \leq -0.7$.
 Therefore, when $|V| \geq 3$, the highest degree vertex will reach all other vertices in 2 hops, which means that each vertex can reach any other vertex within 4 hops. Hence, $d_0 = 4$ and $h=1$.

%
%
%
%

\medskip

The above analysis is based on undirected unweighted graphs.
However, the power law distribution is commonly found in directed graphs by examining the in-degree and out-degree distribution separately \cite{Kunegis12websci,Pareto19Milano}.
The study in \cite{Newman01phyrev} also considers directed graph, and
by focusing on the vertices that can be reached from a random vertex, it is found that many results follow as in undirected graphs. Hence, Assumption \ref{A1} is also for directed graphs.

Based on $d_0$, we have two types of shortest paths: long ones (i.e., those of hop length at least $d_0$) and short ones (i.e., those of hop length below $d_0$).
We have identified hitting sets for covering the long shortest paths based on Assumption \ref{A1}.
Next, we will examine how the shortest paths of hop length shorter than $d_0$ can be handled.

\smallskip

Let $P_{<}$ be the set of all shortest paths $p$ such that $\ell(p) < d_0$, and $P_{\geq}$ be the set of all shortest paths $p$ such that $\ell(p) \geq d_0$. The $d_0$\textbf{-inner-circle} of a vertex $v$ is defined to be $N_<(v) =
 \{p \ | \ {p \in P_{<} \wedge v \in p} \}$. We can visualize $N_<(v)$ as the set of all shortest paths passing through $v$ within a ball with radius $d_0$ centered at $v$, where each path has length less than $d_0$. Similarly, the $d_0$\textbf{-outer-circle} of $v$ is defined as $N_\geq(v) = \{ p \ | \ {p \in P_{\geq} \wedge v \in p} \}$.


\smallskip

We define a neighborhood for vertex $v$ to be used as a hitting set for short shortest paths through $v$.
Let $N(v) = \{u| dist_G(v,u) < d_0 \vee dist_G(u,v) < d_0 \}$,
$N_H(v) = N(v) \cap \mathbb{H}$, and $N''(v) \subseteq N(v)$ be vertices connected to $N_H(v)$ so that for any vertex $u \in N''(v)$, there is a shortest path from $v$ to $u$ or from $u$ to $v$ which contains a vertex in $N_H(v)$.
Then, the set of vertices of $N_e(v) = ((N(v) - N''(v)) \cup N_H(v))$ is called the $\mathbb{H}$-excluded neighborhood of $v$.
If there exists a shortest path $p = v \leadsto u$ with
hop length $< d_0$, then $p$ is hit by a vertex $w$, where $w \in N_e(v)$ and $w \in N_e(u)$.
If we include entries for all vertices in $N_e(v)$ in the label for each vertex $v$, such a shortest path will be found from the labels of the 2 endpoint vertices of the path.
We make an assumption that $N_e(v)$ is small.

\begin{assumption}
In an unweighted scale-free graph $G=(V,E)$, for a vertex $v$,
the $\mathbb{H}$-excluded neighborhood of $v$, $N_e(v)$, contains at most $h$ vertices.
\label{Assumption2}
\end{assumption}


Given an expansion factor of $R$, for a scale free graph
$G = (V,E)$,
$|N_e(v)|$ for $v \in V$ is bounded by $R^{d_0-1}$. If $|V|=1M$, then $R \approx 20$, and if $-0.8 \leq \gamma \leq -0.7$, $d_0 = 4$.
Then, $|N_e(v)| < 20^3 = 8000$.
The actual size of $|N_e(v)|$ is much smaller than this bound since high degree vertices cover a large number of edges in $G$ and their 
expansions are excluded in $N_e(v)$.

The small $h$ value assumption 
is substantiated by our experimental results on a large number of real graphs.
 We say that a graph has \textbf{hub dimension} $h$ if $\forall u \in V, \exists$ a hitting set $H_<$ for $N_<(u)$ such that $|H_<| = O(h)$ and $\exists$ a hitting set $H_\geq$ for $N_\geq(u)$ such that $|H_\geq| = O(h)$. Intuitively, given hub dimension $h$, there exists for each vertex $u$ a set of at most $O(h)$ vertices hitting all shortest paths passing through $u$, which bounds the optimal label size of $u$ by $O(h)$.
%
%
We state our assumption of small hub dimension.

\begin{assumption}  \label{A2}
An unweighted scale-free graph has a small hub dimension $h$.
\end{assumption}



 In summary, we provide realistic assumptions for unweighted directed/undirected scale-free graphs.
Based on Assumption \ref{A2}, the optimal label size is bounded by $O(h)$ for each vertex. 
Our empirical study in Section \ref{sec:exp} shows that for all the scale-free real-world and synthetic graphs that we have tested, the label sizes resulting from our algorithm are very small compared to the graph size. Thus, the assumptions above are strongly supported by experimental results.
The remaining question is how to attain this size bound.

\subsection{Existing Algorithms with Vertex Ordering}

 As discussed in Section \ref{sec:ordering}, ranking of vertices by their degrees has been adopted in PLL \cite{Akiba13sigmod}, and less explicitly in IS-Label \cite{Fu13vldb}. However, as noted in Section \ref{sec:Intro}, both of these methods are not scalable.
For PLL, the in-memory label construction involves many iterations of breadth first search (BFS), and BFS does not yield to an efficient external algorithm to date
\cite{Mehlhorn02esa}.
More importantly, to be efficient, the label pruning in PLL requires a main memory that can hold the labeling index, which is typically much bigger than the given graph.
Hence, it is an open problem to derive an algorithm with scalable bounds on memory and computation consumption and that produces bounded index sizes.
We will focus on this problem for scale-free graphs.

In \cite{Chen12TOA}, it is shown that high-degree vertices in power-law graphs are
useful for finding approximate shortest paths by a compact routing scheme.
A \emph{routing table} is built for each vertex $v$, which keeps track of shortest paths to high-degree vertices called landmarks and to vertices closer to $v$.
However, the query evaluation in \cite{Chen12TOA} does not return exact answers.
In the next sections, we shall make use of vertex degree ordering to derive an I/O efficient algorithm for index construction for exact querying on a large scale-free graph. Our algorithm does not require the knowledge of $h$ but will
seamlessly attain the label size bound of $O(h|V|)$ and scalable complexities. 

%% file: alg.tex

\section{Proposed Solution}
\label{sec:alg}

Our proposed solution is made up of the three major components of algorithmic designs. We first give an outline of each component.
\begin{enumerate}
\item
The basic framework of our label index construction is an iterative process with two steps in each iteration:
 \begin{itemize}
 \item label entry generation based on a set of rules.
 \item label pruning to reduce the label size.
 \end{itemize}
\item
The second design component is an I/O efficient algorithm for implementing the iterative process
(see Section \ref{sec:complexity}).
\item
The third algorithmic design is an enhancement on the performance based on the idea of hop-stepping (see Section \ref{sec:Stepping}).
\end{enumerate}

In this section we describe the iterative process of label generation and pruning.
Based on the discussion in Section \ref{sec:scaleFree},
we design our labeling algorithm with the assumption that
the hitting set of the majority of paths of longer lengths passing through a vertex $v$ is
a small set of $h$ high degree vertices in $\mathbb{H}$.
Since each label entry should correspond to a shortest path, if we place the entries $(v_h,d)$
for vertices $v_h$ in $\mathbb{H}$ in the relevant vertex labels, they would serve most querying.
Analogously, we should try to avoid creating label entries for shortest paths
$p$ = $v \leadsto u$ where $v_h$ is in $p$ for some vertex $v_h \in \mathbb{H}$, and $v_h \not\in \{u,v\}$.
From our assumptions, there are many such paths, and hence many possible label entries, which will lead to large label sizes. 
We will introduce the notion of
\emph{trough paths} for these purposes.

Our strategy is to rank all vertices uniquely according to non-increasing degrees, with the highest rank given to the highest degree vertex.
%
 Next, our algorithm generates label entries to cover shortest paths with increasing number of hops.
There are several reasons for this strategy. Firstly, we need to search the neighborhood of each vertex for the coverage of short shortest paths. Secondly, we need short shortest paths involving $\mathbb{H}$ for pruning other paths. Hence, we traverse from short to long paths.
Thirdly, the iterative approach can be realized by I/O efficient algorithms with scalable I/O complexities, as we will show in Section \ref{sec:Stepping}.
We will explain these points in the following discussion.



\subsection{Iterative Labeling Algorithm}
\label{sec:objectives}

Given a directed unweighted graph $G=(V,E)$,
let $\{v_1, v_2, ..., v_n\}$ be
a ranking of the vertices in $V$
so that the rank of $v_i$, denoted by $r(v_i)$, is equal to $i$.
We rank the vertices in non-increasing order of their vertex degrees. Thus, vertex $v_1$ has the highest degree. We break ties arbitrarily for vertices with the same degree. 
Next we introduce the notion of a trough shortest path.

\vspace*{-2mm}
\begin{definition}[trough shortest path]
A trough path from $v$ to $u$ is a path passing through only vertices with ranks smaller than $\max \{r(u), r(v)\}$. A trough shortest path is a trough path that is also a shortest path.
\end{definition}

\vspace*{-1mm}
For example, in the graph $G$ in Figure \ref{fig:eg} (a), if we rank vertices by non-increasing degrees, then vertex 0 has the highest rank, the
path
$(3,7,2)$ is a trough shortest path, while $(5,3,7)$ is not.
We create labels for each vertex $v$ with the following \emph{labeling objectives}:
\begin{itemize}
\item[\textbf{[O1]}]
if there is a trough shortest path from $v$ to $u$, where $r(u) > r(v)$, then $(u, dist_G(v,u))$ $\in$ $\mathcal{L}_{out}(v)$;
\vspace*{-1mm}
\item[\textbf{[O2]}]
if there is a trough shortest path from $u$ to $v$, where $r(u) > r(v)$, then $(u, dist_G(u,v))$ $\in$ $\mathcal{L}_{in}(v)$.
\end{itemize}


\textbf{Notations:}
Given a label entry $e_1=(u,d_1)$ in $\mathcal{L}_{in}(v)$, it implies that $r(u) > r(v)$ and
there is a trough path $p_1$ from $u$ to $v$ of length $d_1$. $e_1$ is called an \emph{in-label entry}.
We also denote $e_1$ by $(u \rightarrow \underline{v}, d_1)$. 
If there is a label entry $e_2=(v, d_2)$ in $\mathcal{L}_{out}(u)$, it means that $r(v) > r(u)$ and
there is a trough path $p_2$ from $u$ to $v$ of length $d_2$. The entry
$e_2$ is called an \emph{out-label entry}.
We also denote $e_2$ by $(\underline{u} \rightarrow v, d_2)$.
In each case, we say that $e_i$  \textbf{covers} the path $p_i$.
Conversely, given a label entry
$(u \rightarrow \underline{v}, d)$, then
$(u,d)$ $\in$ $\mathcal{L}_{in}(v)$; given
$(\underline{u} \rightarrow v, d)$, then
$(v,d)$ $\in$ $\mathcal{L}_{out}(u)$.
When the ranking is immaterial, we write $(u \to v, d)$, which implies $r(u)>r(v)$ or $r(u)<r(v)$.

\

%% file: alg1.tex

%
%
%
%

In our labeling algorithm, initially each vertex $v$ is assigned two labels $\mathcal{L}_{in}(v) = \{ (v,0) \}$
and $\mathcal{L}_{out}(v) = \{ (v,0) \}$.
In the initialization process, 
for each edge $(u,v) \in E$,
if $r(u)<r(v)$, we add label entry $e = (v,dist_G(u,v)$) to $\mathcal{L}_{out}(u)$;
if $r(u)>r(v)$, we add $e = (u,dist_G(u,v)$) to $\mathcal{L}_{in}(v)$.

\begin{figure}
\center
\includegraphics[height=2.2cm]{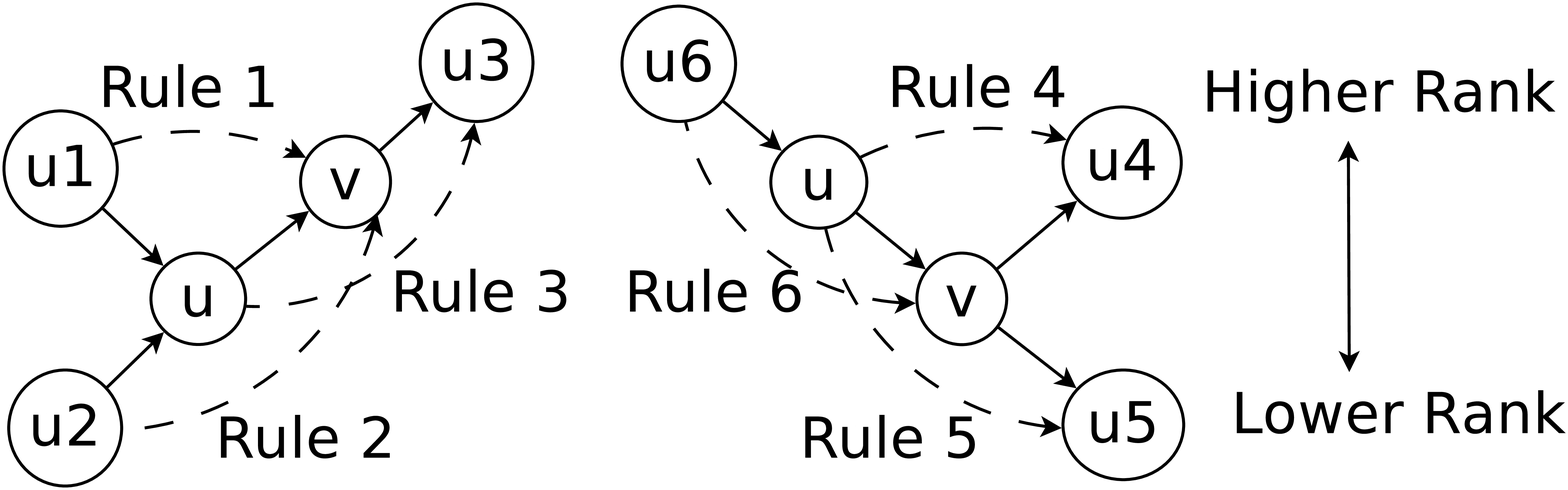}
\vspace*{-3mm}
\caption{Set of label entries generation rules}
\label{fig:1}
\end{figure}

Our algorithm iteratively generates label entries for all vertices until no more label entries can be formed.
The first iteration is the initialization process.
In each remaining iteration, we have a set of new label entries which have been generated in the previous iteration, which we denote by \textbf{prevLabel}. 
Also we have a set of all label entries generated from all previous iterations,
we refer to this set as \textbf{allLabel}.
In each iteration, we adopt 6 rules repeatedly to generate all
the possible label entries for the iteration. The rules are encoded in Table \ref{tab:6rules}. The first rule is derived from the first row in the table as follows:
$\forall (\underline{u} \to v, d) \in prevLabel$, $\forall (u_1\to \underline{u}, d_1) \in allLabel$, generate $(u_1\to v, d_1+d)$. Similarly, the other 5 rules can be derived from the table.
The rules are illustrated in Figure \ref{fig:1}, where each solid or dotted arrow indicates a label entry.

\smallskip


{\small


\begin{table}[h]
\begin{small}
\begin{tabular}{| c| c | c | c |}
  \hline
 & $prevLabel$ & $allLabel$ & $generate$ \\ \hline
 Rule 1 & $(\underline{u} \to v, d)$ & $(u_1\to \underline{u}, d_1)$  &  $(u_1\to v, d_1+d)$ \\
Rule 2 &   $(\underline{u} \to v, d)$ & $(\underline{u_2} \to u, d_2)$ &  $(\underline{u_2}\to v, d_2+d)$ \\
 Rule 3 &  $(\underline{u} \to v, d)$ & $(\underline{v} \to u_3, d_3)$ & $(\underline{u} \to u_3, d_3+d)$\\
 Rule 4 & $(u \to \underline{v}, d)$ & $(\underline{v} \to u_4, d_4)$ & $(u\to u_4, d_4+d)$\\
 Rule 5 & $(u \to \underline{v}, d)$ & $(v \to \underline{u_5}, d_5)$ & $(u\to \underline{u_5}, d_5+d)$ \\
 Rule 6 & $(u \to \underline{v}, d)$ & $(u_6 \to \underline{u}, d_6)$ & $(u_6\to \underline{v}, d_6+d)$ \\
  \hline
\end{tabular}
\vspace*{-3mm}
\caption{Set of label entry generation rules}
\label{tab:6rules}
\end{small}
\end{table}
}


\begin{figure}
\center
\includegraphics[height=2.5cm]{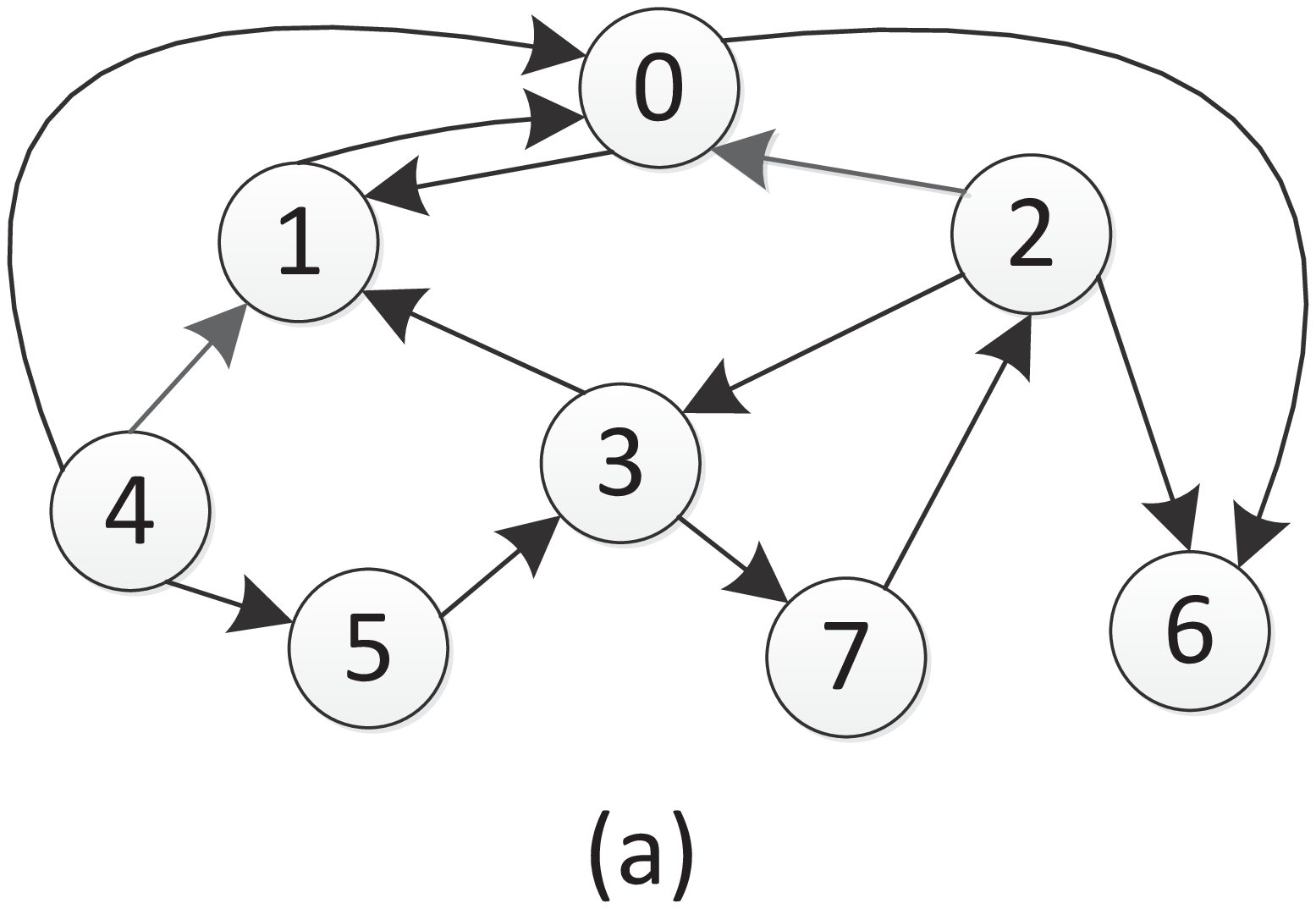}
\hspace*{3mm}
\includegraphics[height=2.5cm]{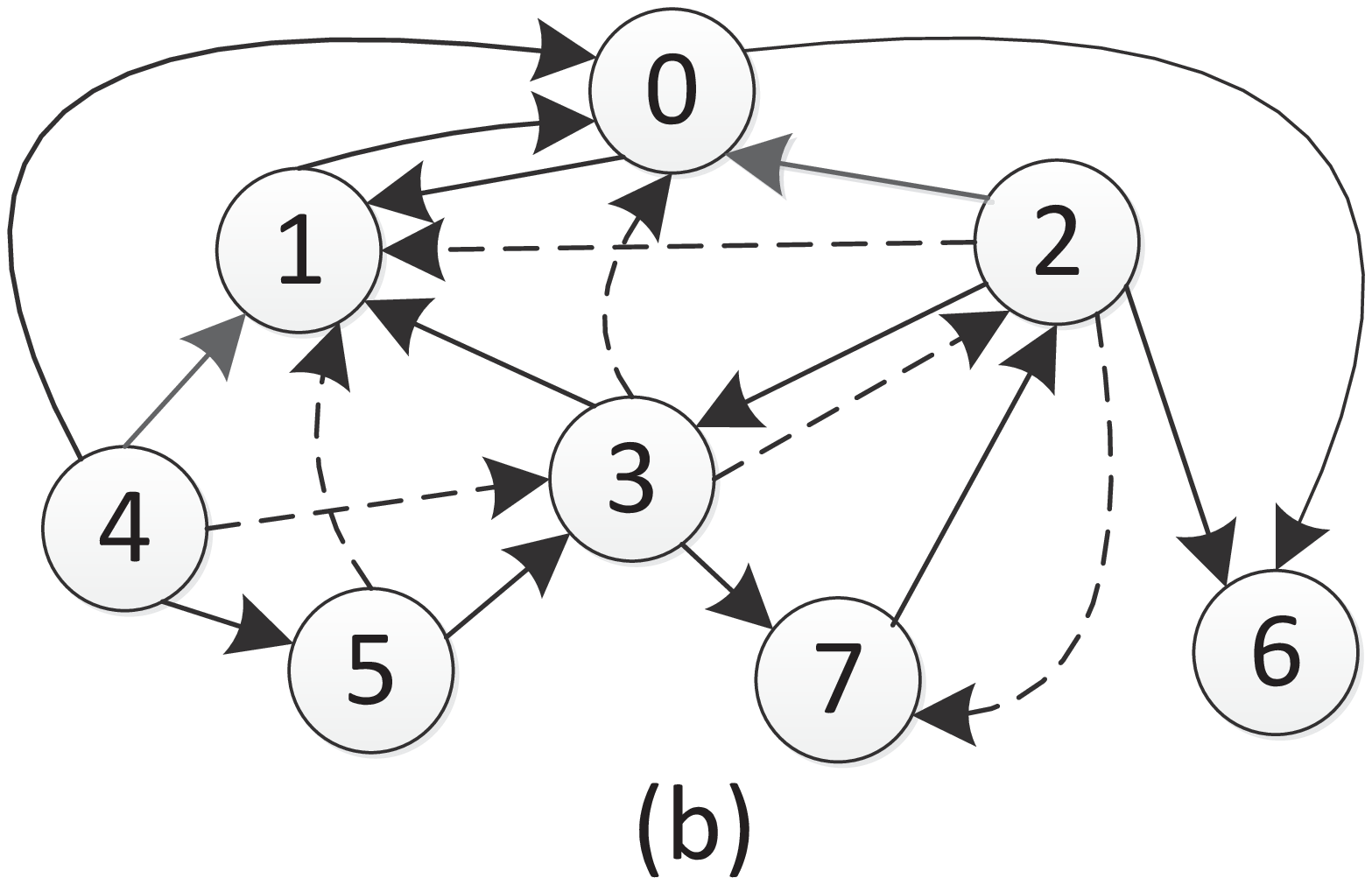}
\vspace*{-3mm}
\caption{(a) Given graph $G=(V,E)$ (b) Trough paths covered after the first iteration (arrows with dotted lines)}
\label{fig:eg}
\end{figure}

A generated label entry $(u \to v, d)$ becomes a new label entry for
the current iteration if there is no existing label entry for $u \to v$, or $d$ is a smaller distance compared with that in other generated or existing label entries for $u \to v$.
When we generate label entry $e$ from two label entries $e_1$ and $e_2$, and given that
$e_1$ covers path $p_1 = (u_1, ..., u_i)$ and $e_2$ covers path $p_2 = (u_i, ..., u_j)$,
then we say that $e$ \textbf{covers} the path $(u_1, ..., u_i, ..., u_j)$.
We shall show that after every two iterations, we double the hop length of trough shortest paths that are covered by the label entries generated. Hence, we call this method \textbf{Hop-Doubling Labeling} (see Algorithm \ref{alg:1}).


\vspace*{-3mm}

\begin{example}
Given the unweighted graph in Figure \ref{fig:eg}(a).
The vertices are ranked by non-increasing degrees and given ID's 0 to 7 accordingly, i.e., vertex 0 has the highest rank. Hop-Doubling Labeling first creates one label entry for
each edge:
$(0 \to \underline{1},1)$, $(\underline{1} \to 0,1)$, $(\underline{2} \to 0,1)$, ...
In the first iteration, by Rule 1 or 4, we generate $(\underline{2} \to 1, 2)$ from
$(2 \to \underline{3}, 1)$ and $\underline{(3 }\to 1, 1)$. Similarly, $(\underline{4} \to 3, 2)$
 and $(\underline{3} \to 2, 2)$ are generated. By Rule 2 or 3, we generate $(\underline{5} \to 1, 2)$ and
  $(\underline{3} \to 0, 2)$, and Rule 5 or 6 generates $(2 \to \underline{7}, 2)$. In the second iteration,
Rule 2 generates $(\underline{4} \to 2, 4)$ from $(\underline{4} \to 3, 2)$ and $(\underline{3} \to 2, 2)$,
Rule 2 also generates $(\underline{5} \to 2, 3)$ and $(\underline{5} \to 0, 3)$. 
In the third iteration,
no new label entry is generated and the labeling is completed. The resulting labels are shown in Figure \ref{tab:eg1}.
\done
\end{example}

\vspace*{-3mm}

\begin{figure}
\vspace*{2mm}
\hspace*{-3mm}
\begin{scriptsize}
\begin{tabular}{|l|l|}
  \hline
  \hspace*{-1mm}
  $\mathcal{L}_{in}(0)$ & \hspace*{-2mm} $\{(0,0)\}$ \\ \hspace*{-1mm}
  $\mathcal{L}_{in}(1)$ & \hspace*{-2mm}  $\{(1,0),(0,1)\}$ \\ \hspace*{-1mm}
  $\mathcal{L}_{in}(2)$ & \hspace*{-2mm} $\{(2,0)\}$ \\ \hspace*{-1mm}
  $\mathcal{L}_{in}(3)$ & \hspace*{-2mm} $\{(3,0),(2,1)\}$\\ \hspace*{-1mm}
  $\mathcal{L}_{in}(4)$ & \hspace*{-2mm} $\{(4,0)\}$ \\ \hspace*{-1mm}
  $\mathcal{L}_{in}(5)$ & \hspace*{-2mm} $\{(5,0),(4,1)\}$ \\ \hspace*{-1mm}
  $\mathcal{L}_{in}(6)$ & \hspace*{-2mm} $\{(6,0),(0,1),$\\
  & \hspace*{-2mm} $(2,1)\}$\\ \hspace*{-1mm}
  $\mathcal{L}_{in}(7)$ & \hspace*{-2mm} $\{(7,0),(3,1),$\\
  & \hspace*{-2mm} $(2,2)^1\}$ \\
  \hline
\end{tabular}
\end{scriptsize}
\hspace*{-2mm}
\begin{scriptsize}
\begin{tabular}{|l|l|}
  \hline
  $\mathcal{L}_{out}(0)$ & \hspace*{-2mm} $\{(0,0)\}$ \\
  $\mathcal{L}_{out}(1)$ & \hspace*{-2mm} $\{(1,0),(0,1)\}$ \hspace*{5mm} \\
  $\mathcal{L}_{out}(2)$ &  \hspace*{-2mm} $\{(2,0),(0,1),(1,2)^1\}$\\
  $\mathcal{L}_{out}(3)$ & \hspace*{-2mm} $\{(3,0),(1,1),(2,2)^1,(0,2)^1\}$\\
  $\mathcal{L}_{out}(4)$ & \hspace*{-2mm} $\{(4,0),(0,1),(1,1),(3,2)^1,$\\
  & \hspace*{-2mm} $(2,4)^2\}$ \\
  $\mathcal{L}_{out}(5)$ & \hspace*{-2mm} $\{(5,0),(3,1),(1,2)^1,(2,3)^2,$\\
  & \hspace*{-2mm} $(0,3)^2\}$\\
  $\mathcal{L}_{out}(6)$ & \hspace*{-2mm} $\{(6,0)\}$ \\
  $\mathcal{L}_{out}(7)$ & \hspace*{-2mm} $\{(7,0),(2,1)\}$\\
  \hline
\end{tabular}
\vspace*{-2mm}
\end{scriptsize}
\caption{Labeling for graph $G$ in Figure 3. The superscript of an entry indicates the iteration in which the entry is generated.}
\vspace*{-3mm}
\label{tab:eg1}
\end{figure}

\vspace*{-4mm}
\begin{algorithm}
\SetKwInOut{input}{Input}\SetKwInOut{output}{Output}
{\small
\input{$G=(V,E)$}
\output{$(\mathcal{L}_{in}, \mathcal{L}_{out})$}
\BlankLine
\tcp{Initialization}
    rank the vertices by non-increasing degrees\;
    $allLabel = prevLabel$ = set of labels covering all edges $e \in E$\;
    \BlankLine
\tcp{iterative construction}
   \While{$prevLabel \neq \emptyset$}{
          Update $prevLabel, all Label$ using the set of label entry generation rules\;
          }
          \BlankLine
build index of $(\mathcal{L}_{in}, \mathcal{L}_{out})$ from $allLabel$\;
}
\caption{Hop-Doubling Labeling}
\label{alg:1}
\end{algorithm}

%% file: properties.tex

Next, we show that distance querying based on the labels constructed by the algorithm is correct. First, we need a lemma.

\begin{lemma}
Hop-Doubling labeling achieves the labeling
objectives of \textbf{[O1]} and \textbf{[O2]} given in Section \ref{sec:objectives}.
\label{lemma1}
\end{lemma}

{\small PROOF}:
Consider a trough shortest path $p$ from $v$ to $u$.
Let the path be $p = ( v = w_1, w_2, ..., w_k = u )$.
We show by induction on the hop length of $P$.
The base case is trivial since we always include $(v,0)$ in
$\mathcal{L}_{in}(v)$ and $\mathcal{L}_{out}(v)$.
Next, assume the statements in \textbf{[O1]} and \textbf{[O2]}
true for all paths of hop length 1 to $k-1$.
Consider the path $p = ( v = w_1, w_2, ..., w_k = u )$.
There are two possible cases.
Case A : $r(w_k) > r(w_1)$;
Case B: $r(w_1) > r(w_k)$.
Let use first consider Case A.
Let $r(w_i)>r(w_j)$ for all
$j < k$ and $j \neq i$.
Since $p$ is a shortest path from $v$ to $u$, the sub-path $p_1$
$= ( w_1, ..., w_i )$ must be a shortest path from
$w_1$ to $w_i$.
Similarly, the sub-path
$p_2 = ( w_i, ..., w_k )$ is a shortest path from
$w_i$ to $w_k$.
Clearly, $dist_G(w_1,w_k) = dist_G(w_1,w_i) + dist_G(w_i,w_k)$.
Since $r(w_i)$ is the second highest rank in $p$, both $p_1$ and $p_2$
are trough shortest paths.
There are two subcases:

Case A1 :  $r(w_i) < r(w_1) < r(w_k)$.
By the induction hypothesis, $e_1 = (w_k, dist_G(w_i,w_k))$ will be
inserted into $\mathcal{L}_{out}(w_i)$, and $e_2 = (w_1, dist_G(w_1, w_i))$ will be
inserted into $\mathcal{L}_{in}(w_i)$.
Note that $e_1 = (\underline{w_i} \rightarrow w_k, dist_G(w_i,w_k))$
and $e_2 = (w_1 \rightarrow \underline{w_i}, dist_G(w_1,w_i))$.
$e_1$ and $e_2$ may be inserted at the same iteration or at
different iterations.
If $e_1$ is inserted in a later round than $e_2$, then
by Rule 1,
$e_3 = (w_k, dist_G(w_1,w_i) + dist_G(w_i,w_k))$ for
$\mathcal{L}_{out}(w_1)$ will be generated.
If $e_2$ is inserted in a later round, then by Rule 4,
$e_3$ will be generated for
$\mathcal{L}_{out}(w_1)$.

Case A2 : $r(w_1) < r(w_i) < r(w_k)$.
By the induction hypothesis,
$e_1 = (w_k, dist_G(w_i,w_k))$
will be inserted into $\mathcal{L}_{out}(w_i)$, and $e_2 = (w_i, dist_G(w_1,w_i))$ will
be inserted into $\mathcal{L}_{out}(w_1)$.
Note that $e_1 = ( \underline{w_i} \rightarrow w_k, dist_G(w_i,w_k))$
and $e_2 = (\underline{w_1} \rightarrow w_i, dist_G(w_1,w_i))$.
If $e_1$ is inserted before $e_2$, then when $e_2$ is newly added,
by Rule 3,
$e_3 = (w_k, dist_G(w_1,w_i) + dist_G(w_i,w_k))$ will be added to
$\mathcal{L}_{out}(w_1)$.
If $e_2$ is inserted before $e_1$, then $e_3$ will be
added to
$\mathcal{L}_{out}(w_1)$ by Rule 2.

Similar arguments hold for Case B 
with subcase B1, where Rules 1 and 4 apply, and subcase B2, where Rules 5 and 6 apply.
\done

\begin{theorem}
The labels constructed by Hop-Doubling Labeling return correct answers for point-to-point distance queries.
\label{thm:correct}
\end{theorem}

{\small PROOF}: By construction, each label entry $(w,d)$ in
$\mathcal{L}_{in}(v)$($\mathcal{L}_{out}(v)$)
covers a path $w \leadsto v$ ($v \leadsto w$) in the graph with length $d$.
Given a distance query from $u$ to $v$,
consider a shortest path $p$ from $u$ to $v$.
Let $w$ be the vertex with the highest rank in $p$.
Note that $w$ can be $u$ or $v$.
Then the sub-paths $u \leadsto w$ and $w \leadsto v$ of $p$ are trough shortest paths.
From Lemma \ref{lemma1} we have an entry $(w,dist_G(u,w))$ in
$\mathcal{L}_{out}(u)$ and also an entry
$(w,dist_G(w,v))$ in $\mathcal{L}_{in}(v)$.
Hence we get the correct distance value of $dist_G(u,v)=$ $dist_G(u,w)+dist_G(w,v)$ when
we look up the labels for $u$ and $v$.
\done

%
%

%% file: 4rules.tex
\subsection{Minimizing the Rules for Labeling}

As illustrated by Figure \ref{fig:1}, we use 6 rules for generating new label entries. In this subsection, we show how to minimize the set of rules to accelerate the generation of new entries. 
For simplicity, here we refer to a label $(u \to v, d)$ as $(u \to v)$.


\begin{lemma}
Rules 1,2,4,5 generate the same results as Rules 1,2,3,4,5,6.
\end{lemma}

\begin{figure}
\center
\includegraphics[width=7.0cm]{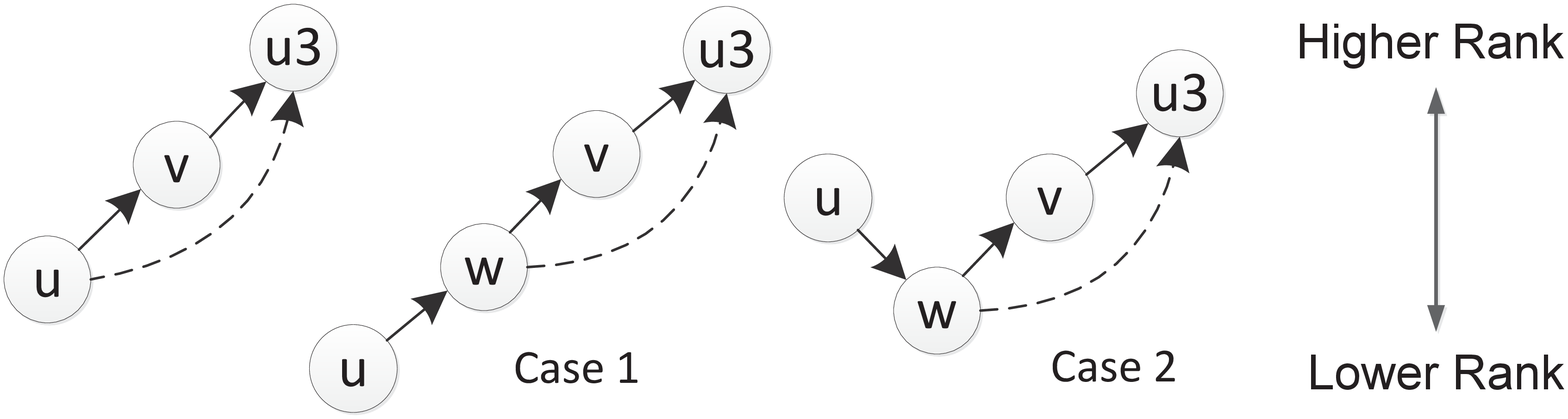}
\vspace*{-3mm}
\caption{Eliminating Rule 3}
\vspace*{-2mm}
\label{fig:3}
\end{figure}

{\small PROOF}: We first prove by induction that label entries generated by Rule 3 can be generated by Rule 1 and Rule 2.
  Assume the lemma holds for all
   iterations up to the $i$-th iteration. At the $(i+1)$-th iteration,
   suppose Rule 3 can generate $(\underline{u} \rightarrow u_3)$ from $(\underline{u} \rightarrow v)$ and $(\underline{v} \rightarrow u_3)$ where
  $(\underline{u} \to v)$ is generated in the $i$-th iteration and
  $(\underline{v} \to u_3)$ is in $allLabel$, then there are two cases of how $(\underline{u} \rightarrow v)$ is generated
   in the $i$-th iteration. (See Figure \ref{fig:3}.)

  Case 1: $(\underline{u} \to v)$ is generated by
  $(\underline{u} \rightarrow w)$ and $(\underline{w} \rightarrow v)$ where $r(u) < r(w) < r(v)$. Hence in the $i$-th iteration, we have $(\underline{u} \rightarrow w)$, $(\underline{w} \rightarrow v)$ and $(\underline{v} \rightarrow u_3)$.
  By Rule 2 we have
  $(\underline{w} \rightarrow u_3)$ before the $(i+1)$-th iteration. Hence, by Rule 2 we can generate $(\underline{u} \rightarrow u_3)$ from $(\underline{u} \rightarrow w)$ and $(\underline{w} \rightarrow u_3)$.

  Case 2 : $(\underline{u} \rightarrow v)$ is generated by $(u \to \underline{w})$ and $(\underline{w} \rightarrow v)$. Hence in the $i$-th iteration, we have
  $(u \rightarrow \underline{w})$, $(\underline{w} \rightarrow v)$ and $(\underline{v} \rightarrow u_3)$. Thus we also have
  $(\underline{w} \rightarrow u_3)$ before the $(i+1)$-th iteration, and by Rule 1 we can generate $(\underline{u} \rightarrow u_3)$ from $(u \rightarrow \underline{w})$ and
  $(\underline{w} \rightarrow u_3)$.

  Thus, $(\underline{u} \rightarrow u_3)$ can be generated in another way with Rule 1 or Rule 2 in the same iteration. Similarly, we can prove that Rule 6 is covered by Rule 4 and Rule 5. 
\done

\begin{figure}
\center
\includegraphics[height=2.0cm]{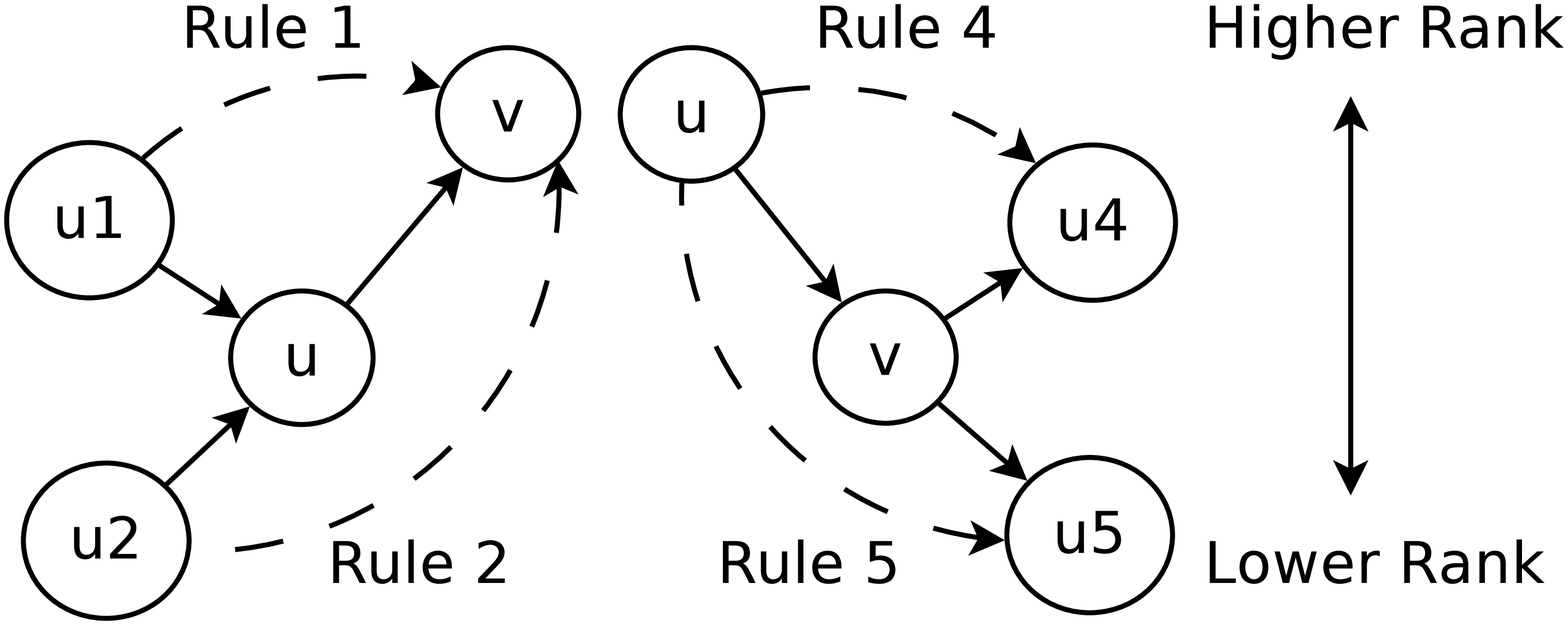}
\vspace*{-2mm}
\caption{4 sufficient rules for label entry generation}
\vspace*{-2mm}
\label{fig:2}
\end{figure}

Other than removing Rules 3 and 6, next, we show that Rules 1 and 4 can be further simplified as follows.
{\small
\begin{enumerate}[leftmargin=0.2in]
\setcounter{enumi}{0}
\item $\forall (\underline{u} \to v, d) \in prevLabel$, $\forall (u_1\to \underline{u}, d_1) \in allLabel$, where $r(v) > r(u_1) > r(u)$, generate $(\underline{u_1}\to v, d_1+d)$
\setcounter{enumi}{3}
\item $\forall (u \to \underline{v}, d) \in prevLabel$, $\forall (\underline{v} \to u_4, d_4) \in allLabel$, where $r(u) > r(u_4) > r(v)$, generate $(u\to \underline{u_4}, d_4+d)$
\end{enumerate}
}

Previously, Rule 1 may also generate ($u_1 \to \underline{v}$), now it only
generates $(\underline{u_1} \to v)$. Similar change applies for Rule 4. The 4 simplified rules are illustrated in Figure \ref{fig:2}.

%

\begin{lemma}
The simplified Rules 1,2,4,5 generate the same results as the original Rules 1,2,4,5.
\end{lemma}

{\small PROOF}:
Consider Rule 1. Originally, we generate $(u_1 \rightarrow v)$ from an old label entry $(u_1 \rightarrow \underline{u})$ and a label entry
 $(\underline{u} \rightarrow v)$ from the previous iteration.
 (1) If $ r(u_1) < r(v)$, then $(\underline{u_1} \to v)$ is also generated 
 by the simplified Rule 1.
  (2) If $ r(u_1) > r(v)$, 
then the label $(\underline{u} \to v)$ must have been generated by either Rule 1 or 2 from
  $(u \to w)$ and $(\underline{w} \to v)$ for some $w$. In the previous iteration or earlier, we
have $(u_1 \to \underline{u})$, $(u \to w)$, and $(\underline{w} \to v)$, by which we also
generate $(u_1 \to \underline{w})$. Then, the simplified Rule 4 will generate $(u_1 \to \underline{v})$.
The arguments for Rule 4 are similar.
\done

%% file: logD.tex
With the above results, the set of rules in
Algorithm \ref{alg:1} now consists of the 4 simplified rules.
We will show that after every 2 iterations, we double the maximum hop length of paths covered by labeling.
 Let $D_H$ be the maximum number of edges among all the pairwise shortest paths. We shall refer to $D_H$ as the \textbf{hop diameter} of the graph.
$D_H$ is the diameter of the graph for an unweighted graph.
We call a path with $k$ hops or edges a \textbf{$k$-length} path.



\vspace*{-1mm}
\begin{theorem}
For all $0 \leq i \leq \lceil \log(D_H) \rceil$,
after the $2i$-th iteration, for each positive integer $k$ $\leq 2^i$, the label entries covering all $k$-length trough paths are generated.
\label{lem:hopDoubles}
\end{theorem}

\vspace*{-1mm}
{\small PROOF}:
We say that a path $p$ is processed if the label entry covering $p$ is generated in the label sets. We prove by induction.
The base case where $i = 0$ is straightforward.
Assume the statement true for $i \leq j$.
We want to show that in the $(2j+2)$-th iteration, the label entries for all $k$-length trough paths are generated where $k \leq 2^{j+1}$.
Consider a $k$-length trough path $p =$$( v_1$,$v_2$,$...,v_{k+1})$,
$k = 2^{j+1}$.
Without loss of generality, assume $r(v_1)< r(v_{k+1})$.
Let $v_j$ be the midpoint of $p$, so that $p$ is divided into 2 paths $p_1 = (v_1 , ..., v_j)$ and
$p_2 = (v_j ,..., v_{k+1})$.
Obviously, $p_2$ is a trough path and it has a hop length of $2^j$, and by induction, its label entry has been generated latest in the $2j$-th iteration.
Let $v_h$ be the vertex of highest rank among $v_1, ...v_j$.
Then, from $p_1$, we have two trough paths $p_{11} = (v_1 ,..., v_h)$ and
$p_{12} = ( v_h ,..., v_j )$.
The hop lengths of $p_1$ and $p_2$ are bounded by $2^j$, and hence both of them are
processed latest in the $2j$-th iteration.
Hence latest at the $(2j+1)$-th iteration, the label entries for the trough path linking $p_{12}$ and $p_2$, i.e. $(v_h, ..., v_{k+1})$ will be created.
Therefore latest at the $(2j+2)$-th iteration, the path $p$ which concatenates $p_{11}$,$p_{12}$ and $p_2$ will be found and processed.
The same argument applies for $ k \leq 2^{j+1}$
\done
%
%
%
%
%
%
%
%
%
%
%
%
%
%
%
%
%
%
%
%
%

%% file: pruning.tex
\vspace*{-2mm}
\subsection{Reducing Index Size by Label Pruning}
\label{pruning}

While the iterative process generates new label entries for trough shortest paths of increasing hop lengths, such a shortest path $p = u \leadsto v$ may be hit by a higher degree vertex $v_h$. We can discover such a case if we find label entries $(\underline{u} \to w,d_1)$ and $(w \to \underline{v},d_2)$, since $w$ is a higher degree vertex.
We add a pruning step in order to remove such generated label entries.
This step is applied to all generated label entries 
at each
iteration after the label generation step at Line 4 of Algorithm \ref{alg:1}.


\vspace*{2mm}

\textbf{Label Pruning:}
A label entry $(u \to v, d)$ is pruned if there exist
label entries
$(\underline{u} \to w, d_1)$
and
$(w \to \underline{v}, d_2)$
where
$d_1 + d_2 \leq d$.


\begin{example}
{
For our example in Figure \ref{fig:eg}, in the first iteration,
$(\underline{2} \to 1, 2)$ is generated from $(2 \to \underline{3},1)$ and $(\underline{3} \to 1,1)$.
However, there exist label entries
$(\underline{2} \to 0, 1)$ and $(0 \to \underline{1},1)$ before this iteration. By the above pruning step,
$(\underline{2} \to 1, 2)$ will be pruned.\done
}
\end{example}

\vspace*{-3mm}

We want to show that with the pruning steps,
the labeling result is still correct.
A similar pruning step is used in PLL \cite{Akiba13sigmod}, but PLL creates label entries by decreasing rank order of the pivots, and thus, the correctness follows from canonical labeling.
It is not obvious in our case since we do not create label entries in rank order.
To show the correctness, we need some definitions.
For the labeling without pruning,
let $L(k)$ be the set of labels at the end of iteration $k$,
and $L$ be the set of labels in the final index.
For the labeling with pruning, let
$L'(k)$ be the set of labels at the end of iteration $k$,
and $L'$ be the set of labels in the final index.

\begin{theorem}[Correctness]
Distance querying by the index built by
Hop-Doubling labeling with pruning is correct.
\label{HDBP}
\end{theorem}

{\small PROOF}:
Given a distance query from $s$ to $t$ in $G$,
consider the set $\mathbb{P}$ of all shortest paths from $s$ to $t$.
Let $p \in \mathbb{P}$ contain the highest ranked vertex $v_m$
in all paths 
in $\mathbb{P}$.
Note that $v_m$ can be $s$ or $t$.
Then, subpaths $(s \leadsto v_m)$ and $(v_m \leadsto t)$ in $p$
are trough shortest paths.
By Lemma \ref{lemma1},
$e_1 = (s \to v_m, dist_G(s,v_m) )$
and $e_2 = (v_m \to t, dist_G(v_m,t))$ are generated in $L$.
We want to show that $e_1$ and $e_2$ are also in $L'$.
We prove by contradiction.
Suppose $e_1$ $\not\in$ $L'$, then
it has been pruned at some iteration $k$, so that $e_1 \in L(k) - L'(k)$.
By the pruning mechanism, at iteration $k$, there exist label entries
$(\underline{s} \to w, d_1)$ and $(w \to \underline{v_m}, d_2)$ from previous iterations,
and $d_1 + d_2 = dist_G(s,v_m)$.
Therefore there exists a path
$(s, ..., w, ..., v_m, ..., t)$
with
a length of $d_1 + d_2 +dist_G(v_m,t) = dist_G(s,v_m) + dist_G(v_m,t)$,
and it is a shortest path from $s$ to $t$.
However, $r(w) > r(v_m)$.
This contradicts our assumption that $v_m$ is the highest ranked vertex
in all shortest paths from $s$ to $t$.
The argument for the case where $e_2$ $\not\in$ $L'$
is similar.
Hence, we conclude that $e_1$ and $e_2$ exist in $L'$ and the answer to the query is correct.\done

\vspace*{-1mm}
\begin{corollary}
Latest at iteration $k = 2 \lceil \log D_H \rceil$, for any shortest path $u \leadsto v$,
there exist the label entries
$(\underline{u} \to v_m, d_1)$ and $(v_m \to \underline{v}, d_2)$ in $L'(k)$ such that $d_1 + d_2 = dist_G(u,v)$.
\label{cor1}
\end{corollary}

\vspace*{-1mm}
The corollary follows from the above proof and Theorem \ref{lem:hopDoubles}, considering that
$v_m$ is the highest ranked vertex among all shortest paths $u \leadsto v$.
Now, we are ready to bound
the number of iterations of our algorithm.

\begin{theorem}
The number of iterations of Hop-Doubling with pruning is upper bounded by
$2 \lceil \log D_H \rceil$.
\label{hopDoubles}
\end{theorem}

{\small PROOF}:
Consider iteration $k = 2 \lceil \log D_H \rceil + 1$, if a label covering a path $p$, $(u \to v, d)$, is generated by one of the 4 rules, then there exists a trough path $u \leadsto v$,
and therefore a shortest path from $u$ to $v$.
From Corollary 1,
there exist in $L'(k-1)$ the label entries
$e_1 = (\underline{u} \to v_m, d_1)$ and $e_2 = (v_m \to \underline{v}, d_2)$ such that $d_1 + d_2 = dist_G(u,v)$,
and these entries will not be pruned in $L'(k)$.
If $v_m = v$, then $(\underline{u} \to v, d_1) \in L'(k-1)$, and $(u \to v, d)$ will not
be generated as a new label. Similarly, if $v_m = u$.
If $v_m \neq v$ and $v_m \neq u$, the label $(u \to v, d)$ will be pruned
by $e_1$ and $e_2$, and will not survive as a new label.
We conclude that no new label will be generated after $2 \lceil \log D_H \rceil$ iterations and
the process stops.
\done

As we shall see in our empirical studies, the above bound is very helpful for some datasets which deviate from the small diameter property of scale-free graphs.

%% file: complexityM2.tex
\section{I/O Efficient Algorithms}
\label{sec:complexity}

In this section, we describe the implementation of Hop-Doubling with pruning
and
analyze the time complexity and I/O complexity.
There are two steps in each iteration:
(1) label generation, which we call \emph{candidate generation} here, and (2) label pruning.
For the analysis of I/O complexity, we adopt the following conventions from
\cite{Aggarwal88CACM}.
Let $scan(N) = \Theta(N/B)$, where $N$ is the amount of data
being read or written from or to disk, $M$ is the
main memory size, and $B$ is the disk block size $( 1 << B \leq M/2)$.

\subsection{Candidate Generation}

We assume that main memory may not be able to hold the label index or even the input graph. Hence we devise an I/O efficient mechanism that resembles a nested loop join for candidate generation.
In the following, for clarity, we refer to a label entry $(u \to v,d)$ as $(u \to v)$. 
In each iteration, we have three types of label entries: \emph{prev entries} are generated in the previous iteration and survived pruning, \emph{candidates} are generated in the current iteration, and \emph{old entries} are all label entries that survived pruning before the current iteration.
Hence, the set of $old$ entries includes the $prev$ entries.

\input{nested}

The pseudo code for candidate generation by Rules 1 and 2 is shown in Algorithm \ref{alg:CandGen}.
We load $prev$ label entries $(\underline{u} \to v)$ and $old$ label entries $(u_1\to \underline{u})$ into memory in the outer loop, which are sorted by $u$ in the corresponding files.
We make sure that for each $u$ where there is a $prev$ out-label entry $(\underline{u}\to v)$, 
we load the $u$ related label entries into memory, i.e. $(\underline{u} \to v), (\underline{u} \to v')$, etc., and
$(u_1 \to \underline{u}), (u_1' \to \underline{u}),$ etc.
Next, we sort all the loaded entries $(u_1 \to \underline{u})$ by $u_1$. Note that the $prev$ entries $(\underline{u}\to v)$ are still sorted by the $u$ values.
In the inner loop, for each $u_2$ where there is an $old$ entry $(\underline{u_2} \to u)$, we load all the $old$ entries starting from $u_2$ into memory, i.e. $(\underline{u_2} \to u), (\underline{u_2} \to u'),$ etc. Candidates are also loaded
in the inner loop block.
After loading the 3 kinds of entries, we generate label entries started from $u_2$ by Rule 1 and Rule 2. For generation by Rule 1, we find $old$ in-label entries $(u_1 \to \underline{u})$
 loaded in the outer loop block with $u_2 = u_1$ by a linear scan of the entries
 $( u_1 \to ...)$. For each $u$, we use a binary search to locate $prev$ out-label entries $(\underline{u} \to v)$, and then enumerate them by a linear scan to
 generate $(\underline{u_2} \to v)$ from $(u_2=u_1 \to \underline{u})$ and $(\underline{u} \to v)$.
We avoid duplicates of $(\underline{u_2} \to v)$ by a binary search among label entries of
$(\underline{u_2} \to ...)$.
For generation by Rule 2, based on $u_2$, we find $prev$ out-label entries $\underline{u} \to v$ to generate $(\underline{u_2} \to v)$ from $(\underline{u_2} \to u)$ and $(\underline{u} \to v)$.
Similarly, we generate candidates from Rules 4 and 5.


Next we analyze the CPU time complexity for candidate generation.
We consider only Rule 1 since the other rules take similar time.
From Theorem \ref{hopDoubles}, there are $O(\log D_H)$ iterations.
In each iteration, for each outer loop block, we scan the $old$ label entries and any candidate label entries generated in
this iteration so far.
Let $|old|$, $|prev|$, and $|cand|$ stand for the total sizes of $old$, $prev$, and candidate label entries,
respectively.
There are $O( (|old|)/M )$ outer loop blocks.
The total CPU time is given by
$O( \log D_H  (|old|)/M  \times |V| |label| \times (\log M + |label|) \times
\log |label|)$,
where $|label|$ bounds the label size of a vertex.
The term $|V|$ comes from each $u_2$ considered in the inner loop block.
For each such $u_2$, we scan $\mathcal{L}_{in}(u_2)$ in the outer block,
thus introducing the factor of $|label|$. For each scanned entry, the binary search and
the linear scan introduce a factor of $(\log M + |label|)$. Finally, $O(\log |label|)$
time is spent for each candidate to avoid duplicates.

For the I/O complexity, we scan $old$ and $prev$ label entries once in the outer loop,
and for each outer loop block, we scan the $old$ and candidate label entries once.
The total I/O cost is thus given by
$O( \log D_H \lceil |old|/M \rceil \times scan(|old|+|cand|) )$.

%

\subsection{Label Pruning}

In each iteration, after the label candidate generation, we apply the pruning step as discussed in Section \ref{pruning}.
For IO efficient computation, we adopt a nested loop join strategy.
We prune an out-label entry $(\underline{u} \to v)$ of $u$ by $(\underline{u} \to w)$ and $(w \to \underline{v})$ where $r(w) > r(v) > r(u)$.
A similar method is adopted for in-label entry $(u \to \underline{v})$ where $r(u) > r(v)$.

 We allocate half of the memory for the outer loop and another half for the inner loop.
In the outer loop, we load $old$ label entries $(\underline{u} \to w), (\underline{u} \to w'), ...$, and candidates $(\underline{u}\to v), (\underline{u} \to v'), ...$, both of which are sorted by $u$, into memory.
In the inner loop, we scan all the $old$ in-label entries $(w \to \underline{v})$,
 $(w' \to \underline{v})$, ..., which are sorted by $v$.
 We scan each $(\underline{u} \to v)$ in the outer loop block.
For each $(\underline{u} \to v)$, we find $v$ related entries $(w \to \underline{v})$ in the inner loop block by a binary search.
Then, we linearly scan the $u$ related entries $(\underline{u} \to w)$ in the outer loop block together with the $v$ related $(w \to \underline{v})$ for possible pruning of $(\underline{u} \to v)$.
After all $(\underline{u} \to v)$ entries are checked, we load another batch of $(w \to \underline{v})$ in the inner loop to check the unpruned $(\underline{u} \to v)$ until all $(w \to \underline{v})$ have been loaded into memory once for pruning all the possible $(\underline{u} \to v)$ in memory from the outer loop.
We continue this process for all the remaining batches of label entries in the outer loop until the end.


We analyze the CPU complexity for the pruning step.
For each candidate or $old$ entry of $(u \to v)$, we perform a binary search and a scanning of
the labels for $u$ and for $v$, hence the time required is
$O(\log D_H (|cand|+|old|)(\log M + |label|))$.
For I/O complexity,
in each iteration, all the $old$ label entries are loaded into memory for $O(\lceil (|cand|+|old|)/M \rceil)$ times, by nested loop.
With $O(\log D_H)$ iterations, it requires
$O(\log D_H (\lceil (|cand| + |old|)/M \rceil \times scan(|old|) + scan(|cand| + |old|))) $ I/Os.


%% file: nested.tex
\begin{algorithm}[!t]

\SetKwInOut{input}{Input}\SetKwInOut{output}{Output}

{\small
\input{$prev, old$ (label entries)}
\output{candidate label entries}

\BlankLine

\tcp{prev $(\underline{u} \to v)$ are sorted by u in file}
\tcp{old $(u_1 \to \underline{u})$ are sorted by u in file}
\tcp{old $(\underline{u_2} \to u)$ are sorted by $u_2$ in file}

    allocate buffer $B_L$ to load next batch of
      $prev$ $(\underline{u} \to v), (\underline{u} \to v')$, ... and $old$ $(u_1 \to \underline{u}), (u_1' \to \underline{u})$, ... , in $B_L$\;
    allocate buffer $B_R$ to load old $(\underline{u_2} \to u), (\underline{u_2} \to u')$... , and candidates $(\underline{u_2} \to u''),(\underline{u_2} \to u''')$..., in $B_R$\;
	\ForEach{block $B_{L}$}{
        sort the $(u_1 \to \underline{u})$ entries in $B_L$ by $u_1$\;
        \ForEach{block $B_R$}{

             \BlankLine
                \tcp{Generation by Rule 1}
                \ForEach{ old $(u_2=u_1 \to \underline{u})$ in $B_L$}{		            	
		            	\ForEach{$prev$ $(\underline{u} \to v)$ in $B_L$}{		
		                	generate candidate $(u_2 \to v) = (u_2=u_1 \to u \to v)$\;
		                }	
		             }
		
		            \BlankLine
		            \tcp{Generation by Rule 2}
\ForEach{$(\underline{u_2} \to u)$ in $B_R$}{
		            \ForEach{$prev$ $(\underline{u} \to v)$ in $B_L$}{
		            	generate candidate $(u_2 \to v) = (u_2 \to u \to v)$\;
		            }
}
%

            }
        }

}
\caption{Candidate Generation (Rules 1 and 2)}
\label{alg:CandGen}
\end{algorithm}

%% file: Stepping.tex
\section{Enhancement by Hop-Stepping}
\label{sec:Stepping}


For Hop-Doubling labeling, the I/O complexity is given by
$O( \log D_H \lceil (|old|+|cand|)/M \rceil \times (|old|+|cand|)/B  )$.
Let us consider $|cand|$.
The candidates are generated from the labels created in the
previous round of execution. From Equation (\ref{R}),
the expansion factor is $R = \log |V|$. In each iteration,
from Theorem \ref{hopDoubles},
the path hop length can expand by at most $D_H/2$, where $D_H$ is the hop diameter of the graph. Hence, $|cand| = O( |prev| (\log |V|)^{D_H/2} )$.
The factor of $(\log |V|)^{D_H/2}$ can greatly affect the I/O cost.
It is caused by the hop doubling property,
where in each iteration we may cover paths with hop lengths up to double that in the previous round.
To deal with this issue,
we consider an alternative strategy 
whereby we increase the number of hops by one
in each iteration.
We show that after each iteration, the label size is bounded by $O(h|V|)$.
Since $R = \log|V|$, the value of $|cand|$ in the complexity analysis becomes
$O(h|V|\log |V|)$.
We call this method \textbf{Hop-Stepping}.

%
%

\subsection{Hop Length $i+1$ from $i$ and 1}

Hop-Stepping retains all the steps of the Hop-Doubling labeling method. However, the 4 rules as illustrated in Figure \ref{fig:2} for generating labels are refined as follows:
at iteration $i+1$,
hop length of the path covered by $u \rightarrow v$ is $i$; while we have
unit hop length for the paths covered by the following labels:
$u_1 \rightarrow \underline{u}$ in Rule 1;
$\underline{u_2} \rightarrow u$ in Rule 2;
$u \rightarrow \underline{u_4}$ in Rule 4;
and
$u \rightarrow \underline{u_5}$ in Rule 5. Only edges in $E$ have unit hop lengths.
E.g., Rule 1 becomes $\forall (\underline{u} \to v, i) \in prevLabel$, $\forall (u_1\to \underline{u}, 1) \in allLabel$, where $(u_1, u) \in E$ and $r(v) > r(u_1) > r(u)$, generate $(\underline{u_1}\to v, i+1)$.

\begin{example}
{
For the graph $G$ in Figure \ref{fig:eg}, in the second iteration of Hop Stepping,
$(\underline{4} \to 2, 4)$ will not be generated, since the hop lengths of
both $(\underline{4} \to 3, 2)$ and $(\underline{3} \to 2, 2)$ are 2.
$(\underline{4} \to 2, 4)$ is generated in the next iteration from
$(4 \to \underline{5}, 1)$ and $(\underline{5} \to 2, 3)$. 
}
\end{example}

%


Let us consider the correctness and other properties of
Hop-Stepping. First, we show that it generates label entries for
paths of unit increasing hop-lengths in subsequent iterations.
In the following, we refer to a path with $i$ hops as an
\textbf{$i$-length} path.

\begin{lemma}
For $1 \leq i \leq D_H$,
at the $i$-th iteration, the
label entries covering all $i$-length trough shortest paths are generated.
\label{hopStepping}
\end{lemma}

{\small PROOF}:
We prove by induction.
The base case where $i = 1$ is straightforward.
Assume the statement true for $1 \leq i \leq j$.
Consider a $(j+1)$-length trough shortest path $p =$ $( v_1$,$v_2$$,...,v_{j+2})$.
Suppose $r(v_1)< r(v_{j+2})$.
$p$ is made up of two sub-paths $p_1 = (v_1, v_2)$ and $p_2 = (v_2, ..., v_{j+2})$.
Obviously $p_2$ is a trough shortest path and it has a hop length of $j$, by induction,
the label entry covering $p_2$ has been generated at the $j$-th iteration.
$p_1 = (v_1, v_2)$ is also a trough shortest path with a hop length of 1,
so the covering entry has also been generated.
By the Hop-Stepping algorithm, $p$ will be generated
at the $(j+1)$-th iteration by either Rule 1 or Rule 2.
Similar arguments hold for $r(v_1) > r(v_{j+2})$
by using Rule 4 and Rule 5. \done

Next, we add the pruning steps to each iteration. We show
that the resulting labeling is correct for distance querying.

\begin{theorem}[Correctness]
Distance querying by the index built by
Hop-Stepping labeling with pruning is correct.
\end{theorem}

The proof is similar to that for Hop-Doubling.
 From Lemma \ref{hopStepping}, we also have the following bound on
 the number of iterations.

\begin{theorem}
The number of iterations of Hop-Stepping labeling with pruning is upper bounded by
$D_H$.
\end{theorem}

\subsection{A Bound on the Label Size}
\label{sec:labelSize}

In this section we derive a bound on the label size.
First we show that after $d_0$ iterations,
only label entries involving vertices in $\mathbb{H}$ (see Assumption 1)
will be added to the labels of each
vertex.

\begin{lemma}
\label{lem7}
Let $l(p) = (u \to v, d)$ be a label
entry which covers trough shortest path $p$,
 where the hop length of $p$ is
$k$ and $k \geq d_0$.
Then, $l(p)$ is pruned at iteration $k$ unless $u \in \mathbb{H}$
or $v \in \mathbb{H}$.
\end{lemma}

\vspace*{-1mm}
{\small PROOF}:
From Lemma \ref{hopStepping}, $l(p)$ is generated at iteration $k$.
Since $p$ has a hop length of $k \geq d_0$, by Assumption \ref{A1},
 $p$ is hit by some vertex in $\mathbb{H}$.
Consider the set $\mathbb{P}$ of all shortest paths from $u$ to $v$ with $k$ hops,
let $w$ be a vertex in $\mathbb{H}$ with the highest rank in $\mathbb{P}$.
Let $h_1$ be the hop length of the shortest path from $u$ to $w$ and
$h_2$ be that from $w$ to $v$. So, $h_1 + h_2 = k$. Hence,
$h_1 \leq k$ and $h_2 \leq k$.
Let us define label sets $L(i)$ and $L'(i)$ as in Section \ref{pruning}.
From Lemma \ref{hopStepping},  $e_1 = (\underline{u} \to w, dist_G(u,w))$ and $e_2 = (w \to \underline{v}, dist_G(w,v))$ are generated at
or before iteration $k$.
We prove by contradiction that
$e_1$ and $e_2$ are in $L'(k)$.
Suppose $e_1 \not\in L'(k)$,
then since it is in $L(k)$, it has been pruned. By
the pruning condition, there exists a higher rank
vertex $x$, with $r(x)>r(w)$, such that $p_2 = (u, ..., x, ..., w)$ has
a length of $dist_G(u,w)$.
Thus, $x$ is a higher ranked vertex that
is on a shortest path from $u$ to $v$, compared to $u$ and $w$, a contradiction to
the fact that $w$ has the highest such rank.
Similarly, we prove that $e_2$ is in the label
of $v$ in $L'(k)$.
Thus, $l(p)$ is pruned at iteration $k$, except
when $w = u$ or $w = v$.
\done

Assumption \ref{Assumption2} in Section \ref{sec:scaleFree}
states that paths of distance below $d_0$ are
hit by a small set of at most $h$ vertices in the close neighborhood if $\mathbb{H}$ is excluded.
Thus, we derive the following.

\begin{lemma}
For each label for each vertex $v$ in the label index $L$, the number of entries
$(u,d)$ where $u \not\in \mathbb{H}$ is bounded by $h$.
\label{lem6}
\end{lemma}

\vspace*{-1mm}

{\small PROOF}: We need only consider $v \not\in \mathbb{H}$ since otherwise $(u,d)$ cannot be in its labels.
$\mathcal{L}_{out}(v)$ initially contains the entries involving out-neighbors of
$v$, then expanding to the close neighborhood with increasing hop lengths.
If no high degree vertex is expanded, this neighborhood is kept small.
Consider a vertex $w \in \mathbb{H}$ in the neighborhood at $k$ hops
from $v$. Thus,
$r(w) > r(v)$. Let the path $p$ from $v$ to $w$ via the $k$ hops be a shortest
path of distance $d_1$.
Consider an out-neighbor $u$ of $w$, where $r(u) < r(w)$, and
$u$ is $k+1$ hops from $v$. Let the path from
$v$ to $u$ via $p$ and $w$ be a shortest path of distance $d_1 + d_2$.
The candidate entry $(u,d)$ will be generated from $p$ and $(w,u)$
with $d = d_1 + d_2$
at the $(k+1)$-th iteration.
From Lemma \ref{hopStepping},
the entries $(v \to w, d_1)$ and $(w \to u, d_2)$ have been generated
in previous iterations since their corresponding hop lengths are less than $k+1$. Candidate $(u,d)$
will be pruned by $(v \to w,d_1)$ and
$(w \to u, d_2)$ since $d_1 + d_2 = d$,
and will not be added to $\mathcal{L}_{out}(v)$.
Similar arguments hold for $\mathcal{L}_{in}(v)$.
The lemma then follows from Assumption \ref{Assumption2} and
Lemma \ref{lem7}.
\done


\vspace*{-2mm}
\begin{theorem}
Given an unweighted scale-free graph $G$, the label size of any vertex at any iteration of
Hop-Stepping with Pruning is $O(h)$.
\label{thm:labelsize}
\end{theorem}

Theorem \ref{thm:labelsize} follows from Lemmas \ref{lem7} and \ref{lem6}, and 
Assumptions \ref{A1} to \ref{A2}. Note that this is an optimal label size if
the value of $h$ is a tight bound on the hitting set size. It is easy to show that Hop-Doubling generates all the label entries that are generated in Hop-Stepping,
and by exhaustive pruning, the label size is the same as that of Hop-Stepping and is bounded by $O(h)$.

\subsection{Complexity Analysis}
\label{sec:stepAnalyse}

The detailed algorithm for Hop-Stepping with Pruning is similar to that
for Hop-Doubling, except that we only consider the $old$ label entries with
only one hop. Thus, the analysis is similar
%
%
to that described in
Section \ref{sec:complexity}, except that we have $D_H$ iterations.
From Theorem \ref{thm:labelsize},
$|old|$ = $|prev|$ = $O(h|V|)$.
Since
$|cand| = |prev| \times R$, where $R = \log |V|$,
$|cand|=O(h|V|\log |V|)$.
Therefore, label generation requires $O( D_H \lceil h|V|/M \rceil \times h \log h |V| \times (\log M + h))$
CPU time and
$O( D_H \lceil h|V|/M \rceil \times scan(h|V|\log |V|)$
I/Os.
%
%
Also, in total label pruning takes
$O( D_H h|V| \log |V| ) (\log M + h)$ CPU time and
$O( D_H \times \lceil h|V| \log |V|/M \rceil \times scan(h|V|)) $ I/Os.

\vspace*{-1mm}
\begin{theorem}
With the assumptions of small $D_H$ and $h$,
the total CPU time for Hop-Stepping with pruning is given by
$O( |V| log M ( |V|/M + log |V|))$,
 and the I/O complexity is
$O( |V| log |V|/M \times |V|/B )$.
\end{theorem}

\subsection{Hop-Stepping and Hop-Doubling}

It is possible to combine the strengths of Hop-Doubling with
that of Hop-Stepping. Hop-Stepping can trim the fast growth of
the lengths of paths covered by label entries at the earlier iterations, when the hop lengths are small.
For graphs where the hop diameter is not very small, a small
fraction of the shortest paths will have long hop lengths. In such a case,
to avoid the larger number of iterations,
we can continue the growth by Hop-Doubling.

\begin{lemma}
If we begin the label construction with Hop-Stepping and switch to
Hop-Doubling after a number of iterations, with the pruning step applied
to all iterations,
distance querying based on
the resulting labeling is correct.
\end{lemma}

%% file: BitParallel.tex
\section{Bit-Parallel Processing}
\label{sec:BP}

In PLL algorithm \cite{Akiba13sigmod}, a bit-parallel approach is introduced to accelerate the memory-based query time for undirected unweighted graph. In this section, we show the method to adapt the bit-parallel scheme to our 2-hop index by a post-processing step.

In bit-parallel approach, there are two kinds of labels in the index for each vertex $v$, namely bit-parallel label $L_{BP}(v)$ and normal labels $L_N(v)$. After generating a 2-hop index $L$ by HopDb, we transform some 2-hop labels into bit-parallel labels $L_{BP}$, and keep the rest as normal labels $L_N$. In a graph $G$, we choose some vertices as roots $R$ from $G$, by default 50 roots. For each root $r \in R$, we select up to 64 $r$'s neighbhours $u$ as $S_r$. Note that $S_r \cap S_{r'} = \emptyset$ if $r \neq r'$. We denote $\bigcup_{r \in R}S_r$ by $S_R$. If a shortest path between $s$ and $t$ can be covered in $L$ as $(s \to u \to t,d_1+d_2)$ by $(u,d_1) \in L(s)$ and $(u,d_2) \in L(t)$, the transformation ensures that it can be covered either by the bit-parallel labels $L_{BP}(s)$ and $L_{BP}(t)$ if $u \in R \cup S_R$, or by the normal labels $L_N(s)$ and $L_N(t)$ if $u \notin R \cup S_R$.

For each vertex $v$, the bit-parallel label $L_{BP}(v)$ is a set of tuples $(r,d_{rv},S^{-1}_r(v),S^0_r(v))$ which stores the distance $d_{rv}$ between vertex $r$ and vertex $v$ for some $r$ with two vertices sets $S^{-1}_r(v)$ and $S^0_r(v)$. A vertex $u \in S^i_r(v)$ implies that $d_{ur}=1$, $d_{uv}-d_{rv}=i$ and $u \in S_r$. To construct $L_{BP}$ from $L$, we initalize $L_{BP}(v) = \{(r,d_{rv},\emptyset,\emptyset) | r \in R, (r,d_{rv}) \in L(v) \}$ for each $v$, and remove such $(r,d_{rv})$ from $L(v)$. We enumerate each $u \in S_r$ for each $r \in R$, and consider each $(u,d_{uv}) \in L(v)$. If $r \notin L_{BP}(v)$, we insert a new tuple $\{(r,d_{rv},\emptyset,\emptyset)\}$ into $L_{BP}(v)$.  From $d_{ur}=1$, we know $d_{uv} - d_{rv} \geq -1$, so if $d_{uv} - d_{rv} = -1$, we insert $u$ into $S^{-1}_r(v)$, or if $d_{uv} - d_{rv} = 0$, we insert $u$ into $S^0_r(v)$. Otherwise, we discard $(u,d_{uv})$ since the path between $v$ and any vertex $x$ via $u$ is impossible to be shorter than than the path via $r$. Then we also remove such $(u,d_{uv})$ from $L(v)$. To query the distance between $s$ and $t$ via $R \cup S_R$, we can check $L_{BP}(s)$ and $L_{BP}(t)$ to find all the common $r$, the distance is $d_{sr}+d_{tr}-2$ if $S^{-1}_r(s) \cap S^{-1}_r(t) \neq \emptyset$, or $d_{sr}+d_{tr}-1$ if $(S^{-1}_r(s) \cap S^{0}_r(t)) \cup (S^{0}_r(s) \cap S^{-1}_r(t)) \neq \emptyset$, or otherwise $d_{sr}+d_{tr}$. 

One way to find all the common $r$ is to take a linear scan on $L_{BP}(s)$ and $L_{BP}(t)$ as checking 2-hop labels, which takes $O(|L_{BP}(s)| + |L_{BP}(t)|)$ time. An optimization is to make use of the property that there are only 50 such $r$. For each vertex $v$, we can use a 50-bit integer as marker $M_{BP}(v)$ to mark the existence of the 50 $r$ in $L_{BP}(v)$. 
With the markers $M_{BP}(s)$ and $M_{BP}(t)$, we can locate the common root $r$ by extracting the 1-value-bit of $M_{BP}(s) \cap M_{BP}(t)$. We also use 50 8-bit integers as offsets for each $v$ to access the label $(r_i,d_{r_iv},S^{-1}_{r_i}(v),S^0_{r_i}(v)) \in L_{BP}(v)$ about $i-th$ root $r_i$ when $r_i$ is found as a common root. Overall, it takes $O(c)$ time to check $L_{BP}(s)$ and $L_{BP}(t)$ if they share $c$ common roots.

To handle the shortest paths via $u \notin R \cup S_R$, we keep the remaining labels in $L$ as normal labels $L_N$. That is to say, for each vertex $v$, the normal label $L_N(v)$, like 2-hop label $L(v)$, is a set of pairs $(u,d_{uv})$ which stores the distance $d_{uv}$ between vertex $v$ and vertex $u$ for some $u$. Checking the shortest paths covered by $L_N(s)$ and $L_N(t)$ is similar to checking 2-hop labels by a linear scan to locate $u \in L_N(s) \cap L_N(t)$. 

%% file: exp3M21.tex

\input{table1MVLDB21}

\input{OtherGraphs1}

\section{Experimental Results}
\label{sec:exp}

We implemented our algorithms in C++, and tested the performance of our algorithms using a Linux machine with an Intel 3.3 GHz CPU, 4GB RAM and 7200 RPM SATA hard disk. To show the advantages of our algorithms, we also compared with three state-of-the-art algorithms, IS-Label \cite{Fu13vldb}, PLL \cite{Akiba13sigmod}, and HCL \cite{JinRXL12sigmod}, with coding provided by their authors. We conducted experiments on various real-world networks. We used a 32-bit integer for each vertex in the vertex set and an 8-bit integer for the distance value in the graph. The information about the datasets is listed in Table \ref{tab:complete}. Most of the datasets are obtained from the Stanford Network Analysis Project and KONECT \cite{KONECT}. We selected graphs with power-law degree distributions. We shall label our algorithm as HopDb.
By default, we adopt the hybrid approach where we apply Hob-Stepping
with pruning in the first 10 iterations and switch to Hob-Doubling with Pruning from the 11-th iteration until the last iteration.

The networks tested in our experiment are as follows.
Delicious is the user-tag network on delicious.com.
BTC is the semantic graph from Billion Triple Challenge 2009.
FlickrLink is the link network on flickr.com.
Skitter is an Internet topology graph.
CatDog and Cat are social networks.
Flickr is the image sharing network on flickr.com.
Enron is an email communication network.
WikiEng/WikiFr/WikiItaly is the wikilinks from Wikipedia.
Baidu is the internal links network on baidu.com.
Gplus and slashdot are social networks.
wikiTalk records the discussions of wikipedia users.
Epinions is a who-trust-who network.
EuAll is a European email network.
AmaRating and EpinRating are customer-product rating networks.
MovRating and BookRating are networks of movie rating and book rating, respectively.
For directed graphs, we rank vertices by non-increasing product of in-degree and out-degree due to its better performance.
%
 We have also considered synthetic scale-free networks generated based on the GLP (Generalized Linear Preference) model \cite{Bu02infocom}. The GLP model is based on the BA model \cite{Barabasi99sci} but allows more flexibility.
The required parameters $m$ and $m_0$ are set to 1.13 and 10, respectively, as in \cite{Bu02infocom}, which gives a
power law exponent of 2.155. Unweighted undirected graphs of varying vertex set sizes and densities are generated,
syn1 to syn6 are six such datasets.

\smallskip
\noindent\textbf{Performance Comparison:}
We compared our algorithm with the only external algorithm IS-Label \cite{Fu13vldb}
which is capable of building full indices.
We also compared our algorithm with the two best existing main memory based indexing methods, namely PLL \cite{Akiba13sigmod} and HCL \cite{JinRXL12sigmod}.
We examined the index size, indexing time, disk based querying time and memory based querying time (with index in memory). Since we are interested in full indexing, we measured the performance of IS-Label when building the complete 2-hop index in Table \ref{tab:complete}. We also compared with baseline bi-Dijkstra search for in memory querying.

The PLL coding provided by the authors of \cite{Akiba13sigmod} only handles undirected unweighted graphs and it incorporated a bit-parallel mechanism for efficient querying, which is applicable to any 2-hop index on undirected unweighted graphs. Hence, we have also added an enhanced bit-parallel component in HopDb for handling the graphs that can be handled by PLL.
The idea of bit-parallel is to select a small set of vertices as roots, e.g. 50 by default in PLL's code, and to merge the label entry of the form $(v,d)$ with $(r,d')$, where $v$ is a neighbour of a root vertex $r$ in the given graph. More details can be found in \cite{Akiba13sigmod}. We also added a bit-wise method to look up common roots in two labels for efficient query processing. 

From the results as shown in Table \ref{tab:complete}, HopDb outperformed the other methods in nearly all aspects.
HCL could not finish all the datasets after running for 24 hours, except for Enron, for which all the costs are 3 orders of magnitude higher than HopDb,
so the results are not included in Table \ref{tab:complete}.
PLL has a smaller indexing time since it is a main memory based algorithm, while HopDB is a disk based algorithm. However, PLL could not handle most of the datasets because of the large main memory requirement for the index construction.
IS-Label could not finish the medium or large sized datasets after running for 24 hours. With the dataset 
Flickr, the intermediate graph $G_i$ has grown to become bigger
than the original graph in the second iteration, and continued to grow.This is because
the pruning strategy of IS-Label 
is much less effective compared with our pruning method.

For the smaller datasets, PLL, IS-Label and HopDb built the complete 2-hop index successfully, but the index sizes of our algorithm are significantly smaller than those of IS-Label and always smaller than PLL, and hence the querying efficiency of HopDb is also substantially better than IS-Label and better than PLL.

We have also conducted experiments on weighted graphs.
While we assume small hitting sets for unweighted graphs only, the results on weighted real graphs also indicate small hitting sets for weighted graphs. This is a promising evidence that the assumptions may also hold for many weighted scale-free graphs.

\medskip

\begin{figure}
\includegraphics[width=6.3cm]{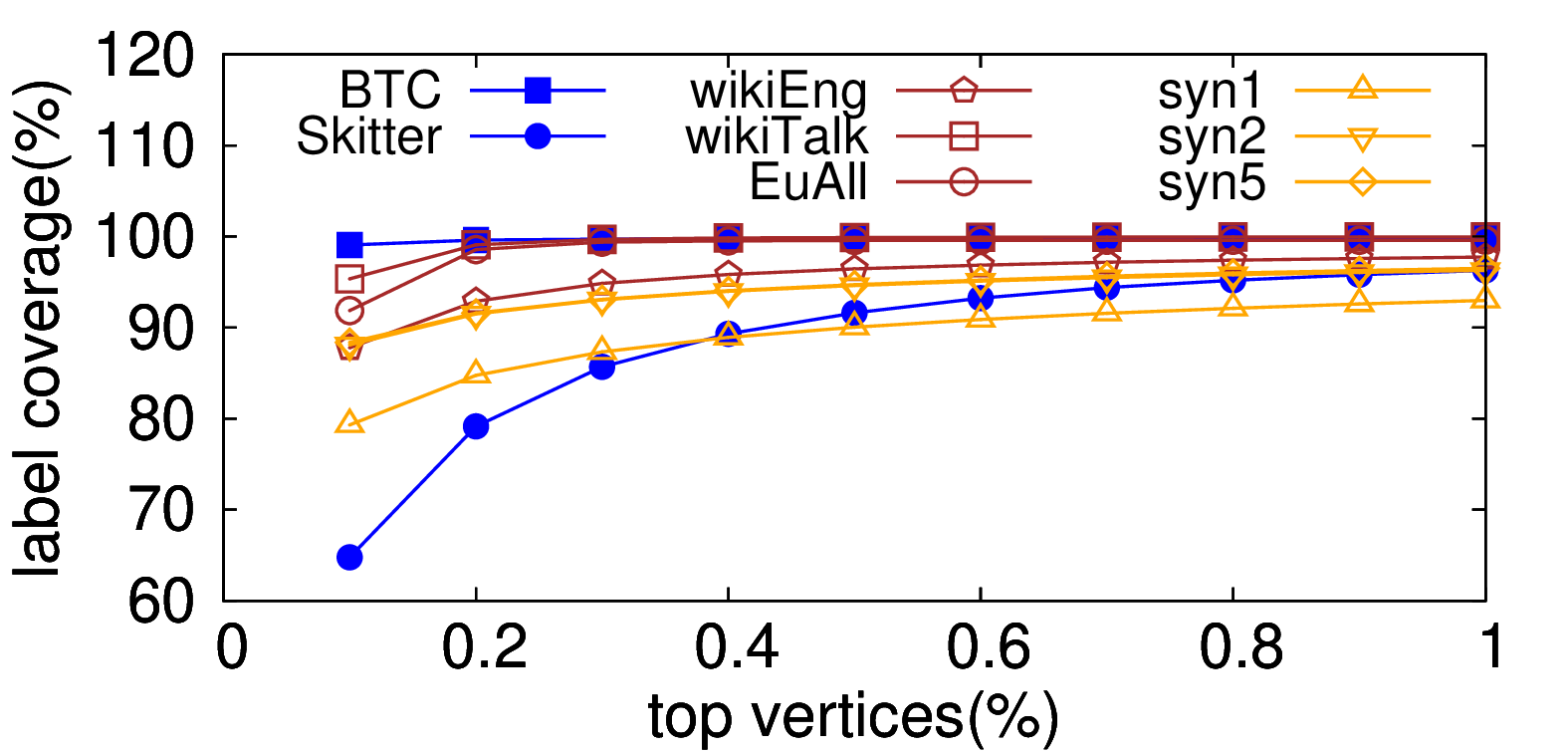}
\vspace*{-2mm}
\caption{Label coverage by top ranked vertices}\vspace*{-2mm}
\label{fig:coverage}
\end{figure}

\noindent\textbf{Results on Small Hitting Set:}
We verify the concept of small hitting set in the real life datasets by showing small average number of label entries ($|label|$) per vertex and high coverage of label entries by top vertices in Table \ref{tab:cover}. 
A label entry $(v,d)$ is said to be covered by $v$.
From our discussion in Section \ref{sec:labelSize}, the size of the final label set can be bounded by $O(h|V|)$ with a small $h$, which is consistent with the small average $|label|$ values listed in the table, and is the guarantee for the high efficiency of our query processing. Moreover, from the label coverage by top vertices, we know that an extremely small amount of top vertices,
given by the percentages in the last three columns of Table \ref{tab:cover},
can cover most label entries, like $70\%, 80\%$, and $90\%$ listed in the table. The top $1\%$ of
vertices often cover close to $100\%$ of the label entries, as shown in
Figure \ref{fig:coverage}. These top vertices formed the set $\mathbb{H}$ for the small hitting sets.

\input{table2M}

\begin{figure}
\includegraphics[height=2.7cm,width=4.1cm]{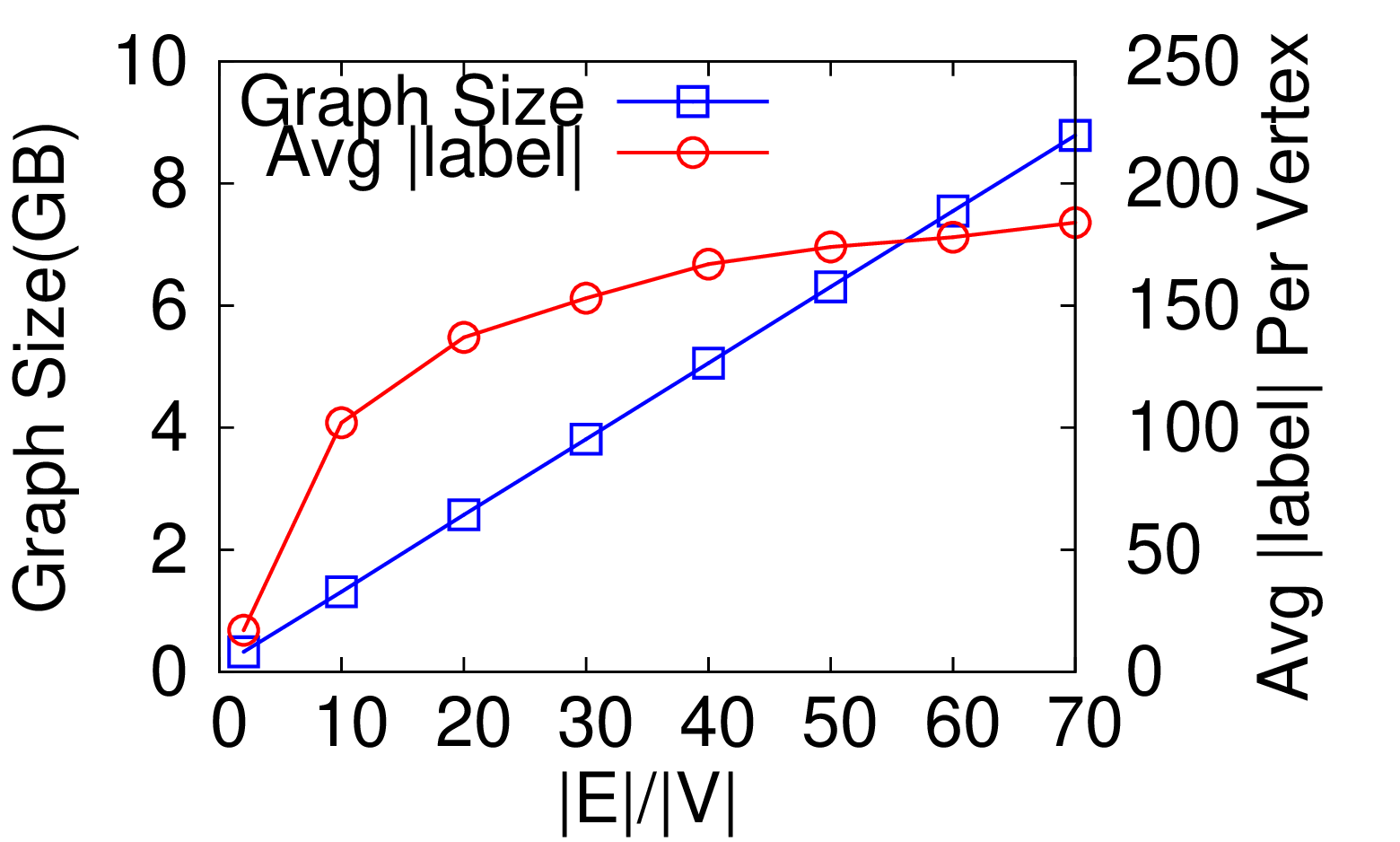}
\includegraphics[height=2.7cm,width=4.1cm]{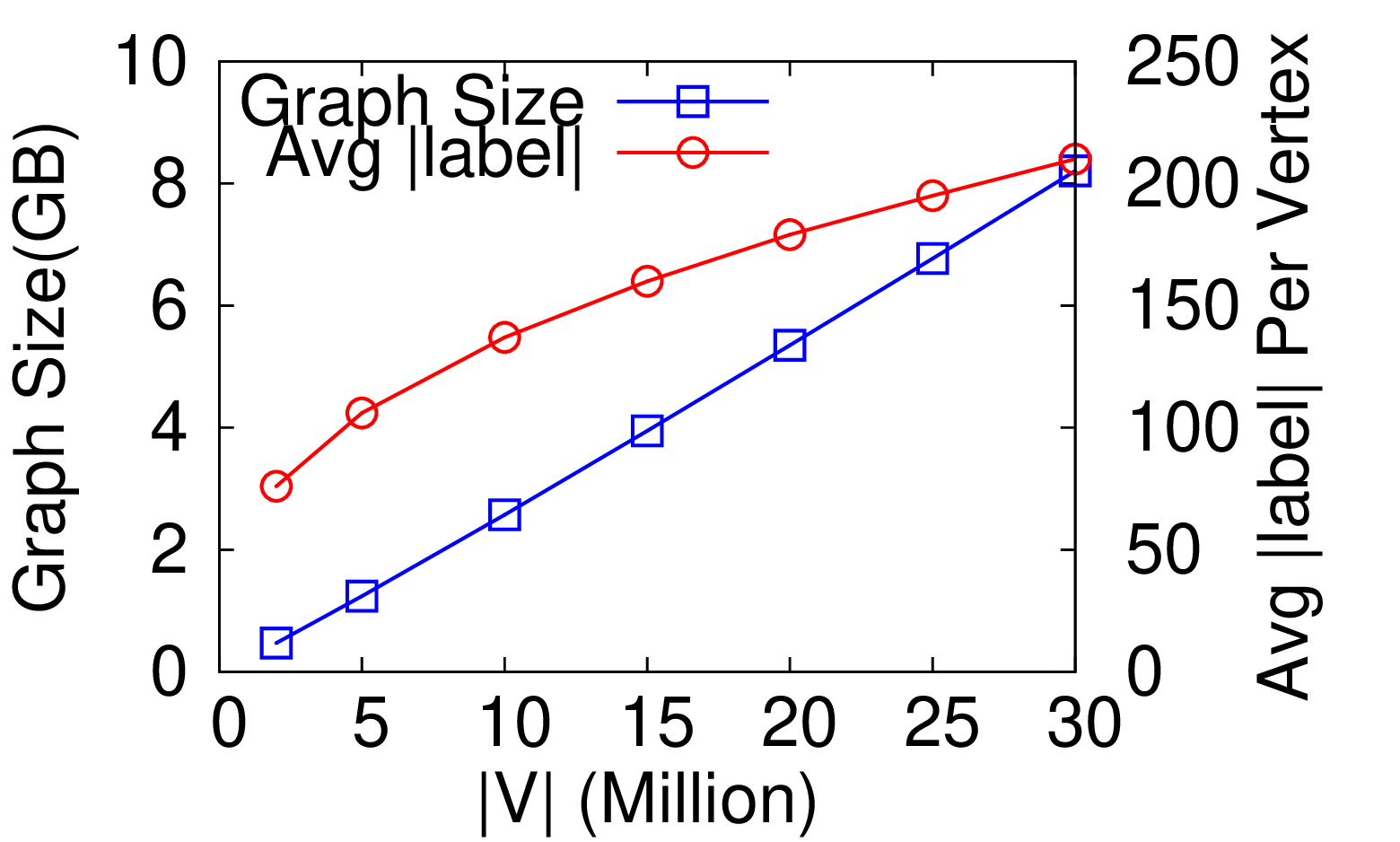}
\vspace*{-4mm}
\hspace*{3mm}(a)\hspace*{40mm}(b)\\
\caption{Results for synthetic scale-free data. (a) $|V| = 10M$ (b) $|E|/|V|=20$}\vspace*{-2mm}
\label{fig:synthetic}
\end{figure}

\medskip

\noindent\textbf{Results on Synthetic Scale-free Data:}
We have generated scale-free networks with different densities in GLP to show the scalability of HopDb. In our first experiment,
the number of vertices of the graphs is fixed to 10 million, and the densities
 $|E|/|V|$ are varied from 2 to 70.
 The number of iterations varies from 7 to 5, which confirms our assumption of a small diameter for scale-free graph.
  The graph sizes and the average number of label entries in a vertex are reported in Figure \ref{fig:synthetic}.
As the graph size increases linearly, the average label size remains very small and approaches a flat value below 200. The
results strongly support our assumptions of small hitting sets and small hub dimension for scale-free graphs.

Similarly, we tested the scalability of HopDb in scale-free networks with growing number of vertices by the GLP model. We set the density $|E|/|V|$ to 20, and varied the number of vertices from 2 millions to 30 millions.
The greatest average label size is around 200, which is very small compared to $|V|$.
This indicates that our assumption of small hub dimension holds for all graph sizes.

\medskip

\begin{figure}[htbp]
\center
\includegraphics[width=6.5cm]{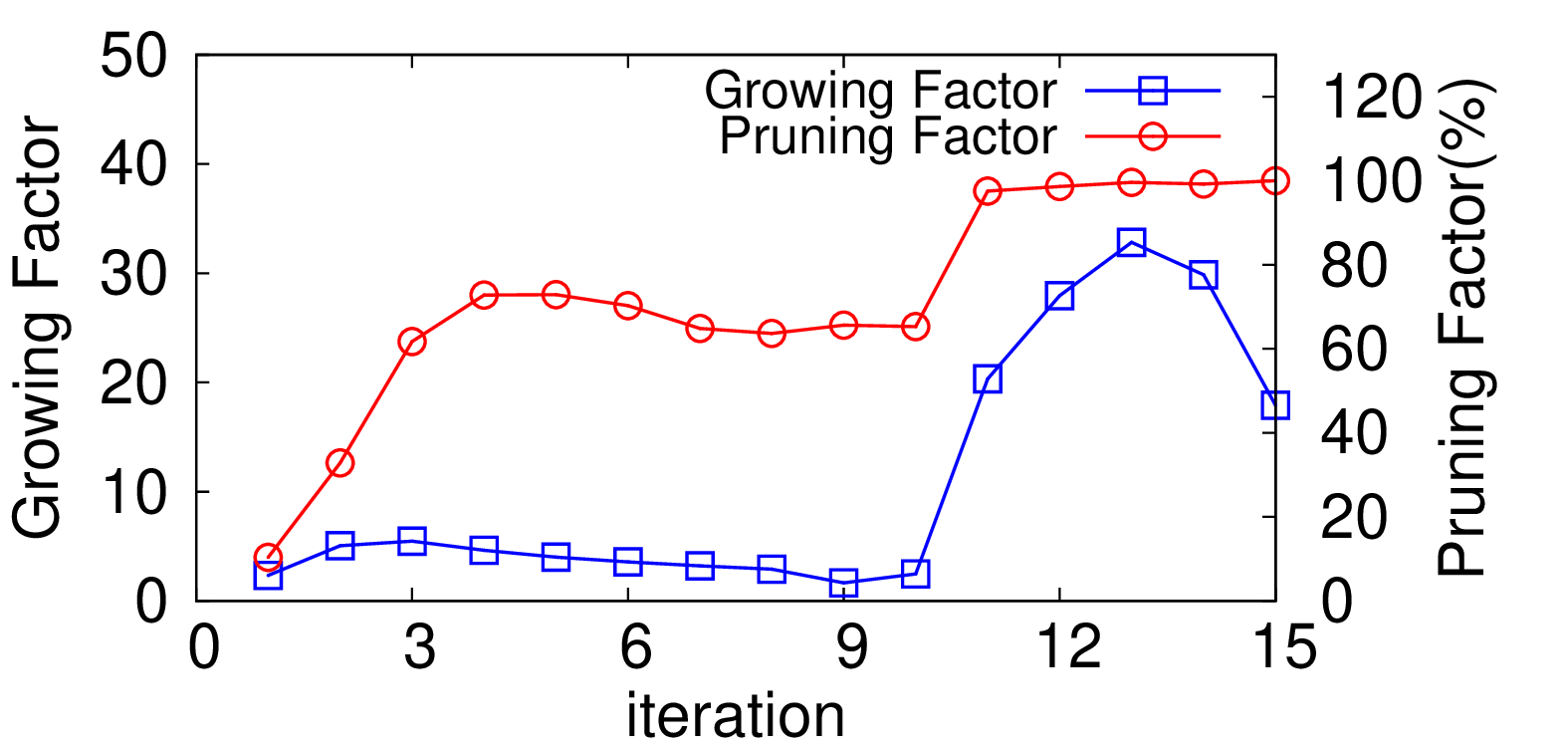}\\
\includegraphics[width=6.5cm]{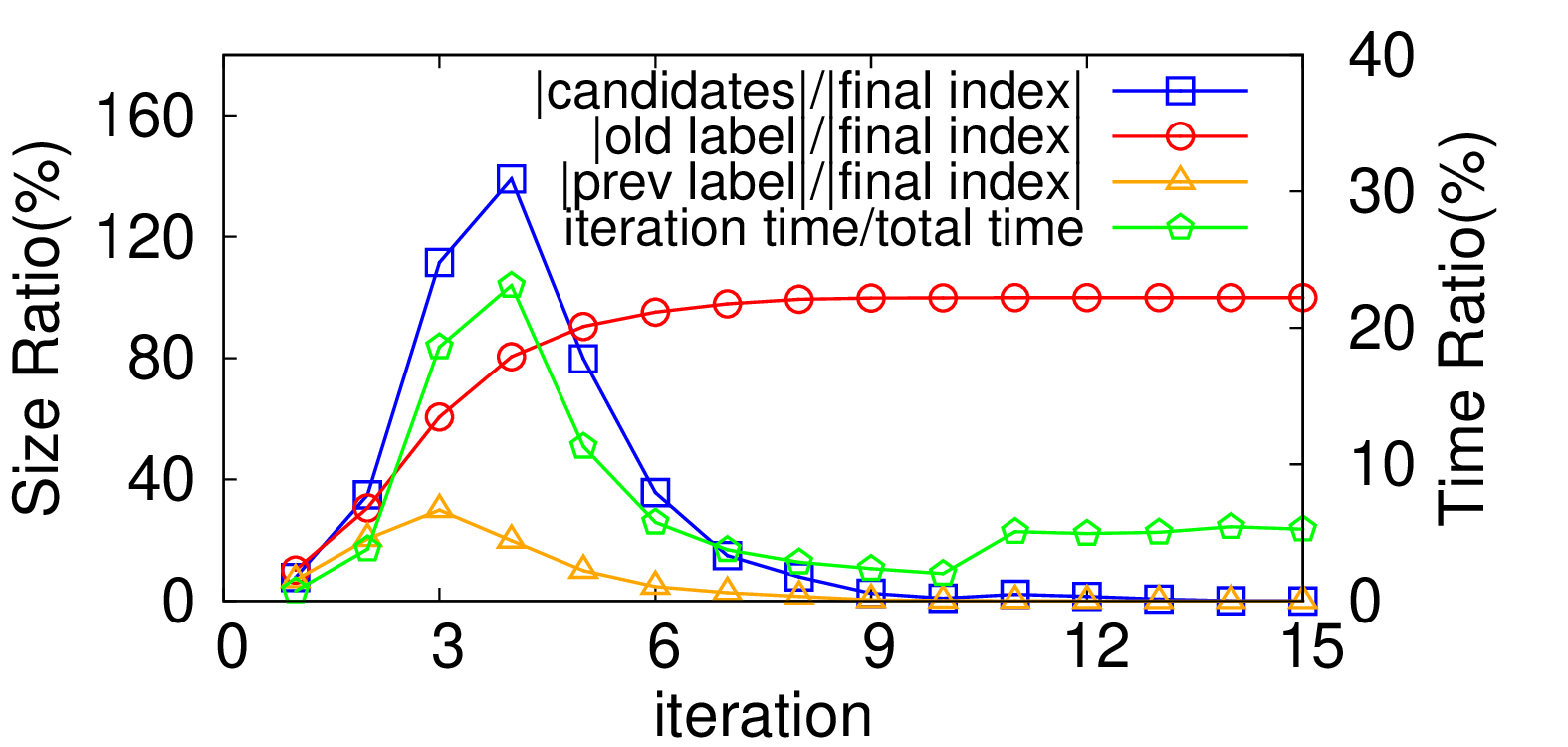}
\vspace*{-3mm}
\caption{Growth and pruning results for wiki-English}
\vspace*{-2mm}
\label{fig:wikiprune}
\end{figure}

\noindent\textbf{Effects of Hop-Stepping and Pruning:}
To show the effectiveness of the
 hop-stepping and pruning strategies, we compared the efficiency of adopting different strategies in Table \ref{tab:DoubleStep} and Figure \ref{fig:wikiprune}.
We considered the three alternatives: only hop-doubling, only hop-stepping, and our default hybrid
approach.
The hybrid approach achieved the best performance as listed in the column hybrid.
Only adopting doubling strategy may lead to too many candidates in the beginning, so it took a long time to finish the large datasets.
In the first 10 iterations, hybrid utilized the hop-stepping strategy to limit the growth of candidates and label size.
From the 11-th iteration, the hybrid approach switched to hop-doubling to accelerate the process of candidate growing and limit the number of iterations.
In datasets with large diameters, 
the hybrid approach could limit the number of iterations and finish the whole process earlier.

\input{table3M}

We analyze the running process of a large dataset, wiki-Eng, to show the power of the pruning strategy and hop-stepping in Figure \ref{fig:wikiprune}. We introduce two numbers, i.e. growing factor and pruning factor, to show the effectiveness.
For each iteration, the growing factor is the ratio of
(\emph{number of candidates generated in this iteration}) to (\emph{number of
label entries generated in the previous iteration}).
%
The pruning factor is the percentage of pruned label entries in one iteration, i.e.
 it is the ratio of (\emph{number of pruned candidate}) to (\emph{total number of candidates}).
The pruning strategy was powerful throughout the whole process. 
In the first 10 iterations when adopting hop-stepping, the growing factor was successfully limited at about 3 to 4, 
this is in line with the small expansion factor
 described in Section \ref{sec:scaleFree}.
After switching to hop-doubling, 
the growing factor
 increased to around 25. Thus, hop-doubling accelerated the label generation and led to earlier termination. In this phase, the effectiveness of the pruning strategy is also shown by the pruning factor, with up to about $90\%$ of the candidates pruned. The runtime of these iterations is short since very few candidates are generated. Figure \ref{fig:wikiprune} also shows that the size of
 the candidate set did not exceed 1.5 times the size of the final index size.
 Hence, the growth in candidates was well under control.

%% file: table1MVLDB21.tex
\begin{table*}
\center
\begin{scriptsize}
\begin{tabular}{ |l|r|r|c|r||r|r|r||r|r|r||r|r|r|r||r|r| }
\hline

\multirow{2}{*}{ $G=(V,E)$}  & \multicolumn{1}{c|}{\multirow{2}{*}{$|V|$}} & \multicolumn{1}{c|}{\multirow{2}{*}{$|E|$}} & \multicolumn{1}{c|}{\multirow{1}{*}{Max}} & \multicolumn{1}{c||}{\multirow{1}{*}{$|G|$}} & \multicolumn{3}{c||}{Index size (MB)}  & \multicolumn{3}{c||}{Indexing time (sec)}& \multicolumn{4}{c||}{Memory query time ($\mu$s)}  & \multicolumn{2}{c|}{Disk query time (ms)} \\ \cline{6-17}

& & & deg ($G$) & (MB)  &  \tiny{IS-Label} & \tiny{PLL} & \tiny{HopDb} &  \tiny{IS-Label} & \tiny{PLL} & \multicolumn{1}{c||}{\tiny{HopDb}}  & \tiny{BIDIJ} &  \tiny{IS-Label} & \tiny{PLL} & \tiny{HopDb} &  \tiny{IS-Label} & \multicolumn{1}{c|}{\tiny{HopDb}}  \\ \hline

\multicolumn{17}{|l|}{undirected unweighted}  \\ \hline

Delicious &  5.3M   &   602M & 4M   & 9446 & --- & --- & 12748& --- &--- & 31999 & ---   & ---  & ---  & ---           & ---  & \textbf{30.1} \\ \hline

BTC       &  168M   &   361M & 106K & 7550 & --- & --- & 13971& --- &--- & 11401 & ---   & ---  & ---  & ---           & ---  & \textbf{28.4} \\ \hline

FlickrLink &  1.7M  &   31M  & 27K  & 452  & --- & --- & 4068 & --- &--- & 4284  & \textbf{25513} & ---  & ---  & ---           & ---  & \textbf{22.7} \\ \hline

Skitter	  &  1.7M	&  22M   & 36K  & 344  & --- & --- & 3732 & --- &--- & 4888  & 5011  & ---  & ---  & \textbf{3.06} & ---  & \textbf{24.6} \\ \hline

CatDog    &  624K	&	16M	 & 81K  & 231  & --- & 836 & 656  & --- &145 & 1152  & 24127 & ---  & 0.98 & \textbf{0.78} & ---  & \textbf{16.3} \\ \hline

Cat       &  150K	&	5M   & 81K  & 67   & 171 & 141 & 61   & 628 & 7  & 102   & 1880  & 2.3  & 0.31 & \textbf{0.22} & 15.7 & \textbf{7.3}  \\ \hline

Flickr    &  106K   &   2M   & 5K   & 30   & --- & 226 & 238  & --- & 42 & 269   & 1497  & ---  & \textbf{2.06} & \textbf{2.06} & ---  & \textbf{12.6} \\ \hline

Enron     &  37K	&	368K & 1K   & 5   & 138 & 33  & 10   & 37  & 0.5& 3     & 108   & 4.8  & 0.14 & \textbf{0.08} & 6.9  & \textbf{0.6}  \\ \hline


\multicolumn{17}{|l|}{directed unweighted}  \\ \hline

wikiEng   & 17M  & 240M  & 2M   & 4447 & ---    &---& 31904 & --- &---& 99686 & --- & --- &---& ---  & ---  & \textbf{38.9}\\ \hline

wikiFr    & 5.1M & 113M  & 1M   & 1964 & ---    &---&  8661 & --- &---& 18532 & \textbf{5317}& --- &---& ---  & ---  & \textbf{31.2} \\ \hline

wikiItaly & 2.9M & 105M  & 825K & 1755 & ---	 &---&	9707 & --- &---& 32397 & \textbf{4384}& --- &---& ---  & ---  & \textbf{28.2} \\ \hline

Baidu     & 2.1M & 18M   & 98K  & 271  & ---    &---&  5184 & --- &---& 6737  & \textbf{1842}& --- &---& ---  & ---  & \textbf{29.4} \\ \hline

gplus	  & 102K & 14M	 & 21K  & 182  & ---	 &---& 337   & --- &---& 623   & 717 & --- &---& \textbf{2.41} & ---  & \textbf{11.6} \\ \hline

wikiTalk  & 2.4M & 5M    & 100K & 74  & ---	 &---& 1464  & --- &---& 377   & 201 & --- &---& \textbf{0.33} & ---  & \textbf{20.4} \\ \hline


slashdot  & 77K  & 517K  & 2K   & 7 &	1035 &---& 65    & 439 &---& 19    & 49  & 7.2 &---& \textbf{0.49} & 18.4 & \textbf{5.7} \\ \hline

epinions  & 76K  & 509K  & 3K   & 6 & 1126   &---& 68    & 517 &---& 20    & 76  & 9.2 &---& \textbf{0.61} & 19.1 & \textbf{4.5} \\ \hline

EuAll     & 265K &	420K & 2K   &  6 &	343	 &---& 65    & 31  &---& 9     & 23  & 8.3 &---& \textbf{0.19} & 11.7 & \textbf{6.3} \\ \hline



\multicolumn{17}{|l|}{synthetic}  \\ \hline

syn1 & 10M  & 700M & 3M & 8998 & ---	 &---  & 9030  & --- &--- &  49612 & ---  & ---  &---& --- & ---  & \textbf{40.1} \\ \hline

syn2 & 20M  & 600M & 4M & 8118 & ---	 &---  & 20272 & --- & ---&  56460 & ---  & ---  &---& --- & ---  & \textbf{37.9} \\ \hline

syn3 & 15M  & 450M & 3M  & 5990 & ---  	 &---  & 13552 & --- & ---&  31920 & ---  & ---  &---& --- & ---  & \textbf{38.2} \\ \hline

syn4 & 10M  & 200M & 2M  & 2633 & ---  	 &---  & 6825  & --- & ---&  7804  & ---  & ---  &---& --- & ---  & \textbf{35.5} \\ \hline

syn5 & 1M   & 5M   & 95K  & 61   & 	7987 & 876 & 161   & 878 & 14 & 43     & 3685 & 40.4 &0.26& \textbf{0.14} & 24.4 & \textbf{15.4} \\ \hline

syn6 & 100K & 1M   & 18K  & 10   & 	262	 & 88  & 14    & 25  &1.4 & 3      & 305  & 3.9  &0.18& \textbf{0.08} & 11.2 & \textbf{1.2} \\ \hline

\multicolumn{17}{|l|}{undirected weighted}  \\ \hline

amaRating & 3.3M & 11M  & 12K & 197 & ---   &---& 15934 & ---  &---& 22609 & \textbf{61450} & ---    &---& ---  & ---   & \textbf{27.7} \\ \hline

epinRating & 876K & 27M & 162K& 376 & ---	&---& 1846  & ---  &---& 2994  & 12550 & ---    &---& \textbf{6.11}  &  ---  & \textbf{22.1} \\ \hline

movRating & 9746 & 2M   & 3K  & 24  & 120	&---& 23    & 452  &---& 50    & 369   & 18.672 &---& \textbf{7.80}  & 4.8   & \textbf{0.8} \\ \hline

bookRating & 264K & 867K & 9K & 13  & 4533  &---& 223   & 2444 &---& 99    & 112   & ---    &---& \textbf{2.28}  & 25.4  & \textbf{14.8} \\ \hline

\end{tabular}
\vspace*{-4mm}
\caption{Performance comparision of BIDIJ, IS-Label, PLL and HopDb on complete 2-hop indexing for different graphs $G$.} 
\label{tab:complete}
\end{scriptsize}
\end{table*}

%% file: OtherGraphs1.tex
\section{Undirected, Weighted, and General Graphs}
\label{sec:othergraphs}

Our algorithms can be easily extended to handle undirected graphs. Instead of having two labels $\mathcal{L}_{in}(v)$ and $\mathcal{L}_{out}(v)$ for each vertex $v$, we need only one label $\mathcal{L}(v)$. To cover an undirected path of length $d$ between $u$ an $v$, where $r(u) < r(v)$, we use the
label entry $(v, d)$ in $\mathcal{L}(u)$.
It is simpler than the directed case, since Rule 1(2) will be identical to Rule 4(5), when the directions of paths are removed. Hence we only need Rules 1 and 2.
For instance, Rule 1 says that: from $(u_1 \to \underline{u},d_1)$ and $(\underline{u} \to v,d)$, where $r(v)>r(u_1)>r(u)$, generate $(\underline{u_1} \to v, d_1 + d)$. For undirected graphs, this rule becomes:
from $(u_1,d_1) \in \mathcal{L}(u)$ and $(v,d) \in \mathcal{L}(u)$, where $r(v)>r(u_1)$,
generate $(v, d_1+d)$ in $\mathcal{L}(u_1)$. Rules 2 is similarly converted.
For distance querying, the labels
$\mathcal{L}(s)$ and $\mathcal{L}(t)$ are looked up for a given query of $dist(s,t)$.

While our discussions so far have focused on unweighted graphs, all our mechanisms also apply to weighted directed/undirected graphs with positive edge weights. 
Though our complexity analysis is based on unweighted scale-free graphs, our experiments on real weighted graphs
show highly promising results.

For graphs that are not scale-free, the ranking by degree may not be effective. For example, road networks do not have high degree vertices. However, our algorithms are still relevant for the general graphs since they work with any total ranking of vertices. As discussed in Section \ref{sec2}, higher ranked vertices should hit a large number of shortest paths. The direct approach to determine such a vertex ranking requires the computation of the shortest paths for all pairs of vertices, which may not be practical for large graphs.
Hence, some heuristical method to approximate this ranking may be helpful. With such a ranking, our algorithms can be applied, and all analyses hold except for those in Sections \ref{sec:labelSize} and \ref{sec:stepAnalyse}, where assumptions based on scale-free graphs are adopted. 

%% file: table2M.tex
\begin{table}
\center
\begin{scriptsize}
\begin{tabular}{ |l||r|r||r|r|r| }
\hline

\multirow{2}{*}{Graph} & number of & Avg $|label|$& \multicolumn{3}{c|}{top vertices coverage}   \\ \cline{4-6}

& Iterations & per vertex & 70\% & 80\% & 90\% \\ \hline

BTC	& 14 & 12 & 0.01\% & 0.01\% & 0.02\% \\ \hline

Skitter	& 13 & 456 & 0.13\% & 0.21\% & 0.43\% \\ \hline

CatDog & 9 & 275 & 0.83\% & 1.55\% & 3.25\% \\ \hline

Cat & 6 & 104 & 0.78\% & 1.33\% & 2.79\% \\ \hline

Flickr & 7 & 515 & 7.62\% & 13.80\% & 16.72\% \\ \hline

Enron & 7 & 321 & 0.60\% & 1.02\% & 2.29\% \\ \hline


%

wikiEng & 15 & 192 & 0.03\% & 0.05\% & 0.13\% \\ \hline

wikiItaly & 15 & 343 & 1.69\% & 2.34\% & 3.73\%  \\ \hline

gplus & 8 & 342 & 2.87\% & 4.37\% & 7.56\%  \\ \hline

wikiTalk & 7 & 60 & 0.02\% & 0.04\% & 0.07\%  \\ \hline


slashdot & 9 & 84 & 0.73\% & 1.12\% & 1.89\% \\ \hline

epinions & 9 & 91 & 0.89\% & 1.31\% & 2.10\% \\ \hline

EuAll & 7 & 22 & 0.04\% & 0.06\% & 0.09\% \\ \hline

%

\end{tabular}
\end{scriptsize}
\vspace*{-2mm}
\caption{Results supporting the assumptions of small hub dimension $h$ and small hitting sets ($|label|$ = number of label entries)}
\vspace*{-2mm}
\label{tab:cover}
\end{table}

%% file: table3M.tex
\begin{table}
\begin{scriptsize}
\begin{center}
\begin{tabular}{ |l||r|r|r||r|r|r| }
\hline

\multirow{2}{*}{Graph}   & \multicolumn{3}{c||}{Indexing time (sec)}  &  \multicolumn{3}{c|}{number of iterations}  \\ \cline{2-7}

 & Double & Step & Hybrid & Double & Step & Hybrid \\ \hline

BTC & --- & 21081 & 11401 & --- & 38 & 14 \\ \hline

Skitter & --- & 6400 & 4888 & --- & 21 & 13 \\ \hline

wikiItaly & --- & 47558 & 32397 & --- & 59 & 15 \\ \hline

gplus & 4205 & 642 & 642 & 5 & 8 & 8 \\ \hline

wikiTalk & 2221 & 378 & 378 & 5 & 7 & 7 \\ \hline

slashdot & 145 & 19 & 19 & 5 & 9 & 9 \\ \hline

epinions & 157 & 20 & 20 & 5 & 9 & 9 \\ \hline



\end{tabular}
\end{center}
\vspace*{-4mm}
\caption{Comparing Hop-Doubling, Hop-Stepping, and Hybrid}
\vspace*{-2mm}
\label{tab:DoubleStep}
\end{scriptsize}
\end{table}

%% file: conclusion.tex
\section{Conclusion}
\label{concl}

We introduce a new disk-based indexing algorithm for distance querying on a large scale-free graph.
The design is based on
properties of unweighted scale-free graphs. 
With scalable indexing complexities,
our method performs well on different types of
scale-free networks and can handle graphs many times
larger than existing methods. The consistently small label sizes resulting from our label indexing
with all our tested graphs strongly support our assumption of small hub dimension. The experimental result also verifies the scalability of our algorithm and
the small label sizes give rise to highly efficient query 
evaluation both in-memory and on-disk.